\documentclass[reprint, amsmath,amssymb, aps, onecolumn]{revtex4-2}
\usepackage[utf8]{inputenc}  
\usepackage{graphicx}
\usepackage{bm}
\usepackage{amsmath}
\usepackage{url}
\usepackage{xcolor}

\usepackage{here}

\usepackage{algorithmic}
\usepackage{algorithm}

\usepackage{subcaption}

\usepackage{appendix}

\PassOptionsToPackage{colorlinks=true, linkcolor=blue, citecolor=blue, urlcolor=blue}{hyperref}

\usepackage{CJKutf8}

\let\doi\relax
\usepackage{doi}

\begin{document}

\title{A Model of Silence, or the Probability of \textbf{\textit{Un Ange Passe}}}
\author{Keishu Utimula}
\date{\today}

\begin{abstract}

In French, the phrase {\it Un ange passe} (``An angel passes'') refers to the sudden silence that falls over a co-present group — that is, a group of people sharing the same physical space.
As evidenced by the presence of similar expressions across languages and cultures, this phenomenon represents a universal feature of human conversation.
At the same time, the meaning attributed to silence can differ greatly across national, cultural, and interpersonal contexts.
Consequently, a wide range of studies have focused on the impact of silence on organizational productivity, its relationship to ideas and creativity, and its potential effectiveness in medical settings.
Despite the important role that silence plays, very few studies have attempted to characterize its features using mathematical modeling. 
In this study, we propose a Markov chain model to describe the dynamics of silence in a co-present group and attempt to analyze its behavior.
Our results reveal a phase-transition-like phenomenon, where the probability of silence abruptly drops to zero once individuals' awareness of the surrounding conversation falls below a critical threshold. 
In other words, such silence can emerge only when individuals retain a minimal degree of mutual awareness of those around them. 
The model proposed in this study not only offers a deeper understanding of conversational dynamics, but also holds potential for contributing to intercultural communication, organizational productivity, and medical practice.

\end{abstract}

\maketitle

\section{Introduction}
\label{sec.intro}

The French phrase {\it Un ange passe}, literally meaning ``an angel passes,'' refers to the sudden silence that occurs when a conversation unexpectedly comes to a halt.
Similar expressions can also be found in other languages, such as the German {\it Ein Engel geht durchs Zimmer} (``an angel passes through the room'') and the English {\it An angel is passing}. A much earlier reference appears in ancient Greece, in Plutarch's observation that ``when in some gathering silence suddenly descends, they say that Hermes has entered the room''~\cite{2000BET}.
In Dutch, the phrase {\it Er gaat een dominee voorbij} (``a pastor passes by''), and in Russian, \raisebox{-0.8ex}{\includegraphics[height=1.4em]{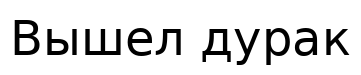}} (``a fool has been born''), are also sometimes used as idiomatic expressions to describe moments of sudden silence.
Thus, the existence of similar expressions across many different languages suggests that the sudden silence occurring within a co-present group may be a universal phenomenon of human interaction.

\vspace{2mm}
Silence can carry meaning beyond the mere absence of speech, and its interpretation and perceived value vary across cultures~\cite{2023TAN, 1986TAN, 2012NAK, 2013JAW, 2023JON, 2012STE, 2013MOL, 2023WAI, 2024YAN}. 
For instance, a study analyzing the conversational style of East European Jews in New York showed that they tend to evaluate silence negatively and therefore strive to maintain a continuous conversational flow~\cite{2023TAN}.
In contrast, silence within Western Apache society reflects the instability of interpersonal relationships~\cite{1970BAS}. 
This depends on the social context: for example, in initial encounters, silence functions as a gesture of respect toward the other person's social status, while in the early stages of courtship among young people, it is perceived as an expression of nervousness or shyness~\cite{1970BAS}.
Furthermore, in reunions between parents and children, silence may signal anxiety or emotional distance, while in situations of conflict, it can function as a means of self-protection~\cite{1970BAS}.
Moreover, in mourning, silence serves as a means of expressing and sharing grief~\cite{1970BAS}.
In Japan, while silence can sometimes be intended as a form of rejection, it is also associated with positive connotations, as reflected in the proverb “Speech is silver, silence is golden”~\cite{2023JON}.
It is used as a means of avoiding conflict and maintaining harmony~\cite{2023JON}. 
Among African communities, silence often plays an important role as part of communication~\cite{1994PEE}. For instance, among the Akan people of West Africa, silence can signify respect for social status and age~\cite{2002AGY}.

\vspace{2mm}
Such differences in the perception of silence can be observed even within a single country. 
A study conducted at three schools in Wales revealed that students in rural and inner-city areas valued silence more highly than those in suburban schools. Furthermore, it pointed out a power imbalance in the perception of silence between students and teachers~\cite{1998JAW}.
Moreover, the interpretation of silence also depends on context~\cite{2011KUR, 2021CUR, 2023TAN2}, and some studies have proposed frameworks for accommodating its polysemous nature~\cite{2023TAN2}.
Therefore, the polysemous nature of silence and its cultural variations may lead to serious misunderstandings, particularly in intercultural communication~\cite{1986TAN, 2012NAK, 2013JAW}. 

\vspace{2mm}
Numerous studies have examined the relationship between silence and productivity. 
A study focusing on the importance of silence in workplace conversations proposed a framework for discussing how communication contributes to productivity, and examined how silence influences productive dialogue~\cite{2021EDM}.
Silence has also been shown to influence individuals' perception of time and self-awareness, and in some cases, it can induce a deep state of relaxation~\cite{2020PFE}.
Notably, one study demonstrated that silence within organizations can reduce employee stress, enhance overall productivity, and improve decision-making processes~\cite{2024ASS}.
In addition, several studies have examined the relationship between silence and creativity. One study suggests that silence conducive to daydreaming is essential for cultivating creative introspection~\cite{2012SAU}, while another argues that silence not only provides a space for generating new ideas and creativity, but also serves as a means to deepen interpersonal relationships~\cite{2018BIG}.

\vspace{2mm}
The significance of silence in medical settings has also been increasingly recognized~\cite{2016HAS, 2022KAY}. 
For example, one study has pointed out that compassionate silence can help foster mutual understanding between patients and healthcare providers, and contribute to building a warm, supportive relationship~\cite{2009BAC}.
In counseling, some perspectives suggest that a client’s silence constitutes an important process through which individuals are encouraged to confront their inner selves~\cite{2009ITO}.
A study analyzing recorded conversations between doctors and patients introduced the concept of ``connectional silence'' to investigate how different types of silence affect the quality of communication~\cite{2016BAR}. 

\vspace{2mm}
As demonstrated above, silence not only possesses culturally significant characteristics, but also has practical implications for organizational productivity and the quality of healthcare.
Despite its significance, only a limited number of studies have attempted to analyze the behavior of silence using models such as mathematical frameworks. 

\vspace{2mm}
An early study that approached human conversation through mathematical modeling and examined silence is that of H. B. Kekre {\it et al}.~\cite{1977KEK}.
They introduced a two-state Markov model alternating between talkspurts (speaking periods) and pauses (silent periods), and analyzed the statistical distribution of these states~\cite{1977KEK}. 
Rather than a genuine interest in silence itself within human conversation, this study was primarily motivated by engineering needs to optimize the use of limited communication bandwidth at the time~\cite{1977KEK}. 
Since the 2000s, for example, O. I. Sheluhin {\it et al.} have proposed using Markov models to improve the efficiency of communication systems in telephone speech dialogues~\cite{2012SHE}. 
In addition, M. Galley {\it et al.} addressed the problem of segmenting multi-party conversations, such as meetings, and developed a method that integrates lexical and conversational features~\cite{2023GAL}.
Notably, in this approach to characterizing conversation, silence—previously overlooked in many studies—plays a significant role~\cite{2023GAL}.
While these studies have highlighted important aspects of silence, their primary focus was not on silence itself. In particular, the qualitative and quantitative behavior of silence in co-present groups remains largely unexplored. 

\vspace{2mm}
On the other hand, the analytical approach to conversation is often referred to as Conversation Analysis (CA), and has been the subject of extensive research~\cite{2023HOE}. 
Conversational Exchange Analysis (CEA), proposed by A. P. Thomas {\it et al.}, codes conversation at three distinct conceptual levels—Activity, Type, and Focus—and provides rules for segmenting and categorizing speech~\cite{1982THO}.
M. Ishizaki and T. Kato quantitatively investigated the characteristics of triadic conversations and found that dyadic interactions tend to be dominant~\cite{2023ISH}.
A. Stolcke {\it et al.} introduced a methodology for analyzing the dynamics of dialogue structure by integrating n-gram models with hidden Markov models~\cite{2000STO}.
P. M. Aoki {\it et al.} used machine learning to analyze participants’ turn-taking behavior and dynamically identify conversational floors, enabling customized audio delivery for each participant~\cite{2023AOK}.
Influence models have also been proposed to model human interactions through conversation~\cite{2001BAS, 2007DON}. 
K. Otsuka {\it et al.} estimated conversational structures using a dynamic Bayesian network known as the Markov-switching model~\cite{2005OTS}.
Of particular note is that they proposed a probabilistic framework for modeling the relationship between conversation and nonverbal behavior by incorporating information such as gaze patterns and head orientations~\cite{2005OTS}.
M. Fujimoto and I. Daido applied principal component analysis (PCA) to multi-party conversations to investigate how individual speaking behavior influences the depth and flow of conversation~\cite{2007FUJ}.
The Conversation Clock method, proposed by T. Bergstrom and K. Karahalios, enables the visualization of conversation patterns based on participants’ aural contributions and allows users to reflect on the dynamics and flow of the conversation~\cite{2007BER}. 
M. Mastrangeli {\it et al.} developed an agent-based conversation model and investigated the relationship between conversational dynamics in multi-party settings and participants’ levels of happiness~\cite{2010MAS}.
M. Cristani {\it et al.} integrated a Gaussian mixture model with a Markovian influence model and constructed a method for classifying dialogue scenarios~\cite{2010CRI}. 
D. Angus {\it et al.} introduced the conceptual recurrence plot technique to assess the conceptual similarity between arbitrary pairs of utterances, going beyond term-based approaches~\cite{2012ANG}.
They applied this approach to doctor–patient dialogues and explored the relationship between communication patterns and trust~\cite{2012ANG2}.
R. Fusaroli and K. Tylen analyzed dialogues using recurrence quantification analysis (RQA) and demonstrated that synergy serves as a strong predictor of group performance~\cite{2015FUS}.
L. A. Clarfeld {\it et al.} developed a new approach, called the COnversational DYnamics Model (CODYM), which provides a unified framework for assessing conversational dynamics and the flow of information~\cite{2020CLA}.
This model is a Markov model that uses speaker turn lengths as its fundamental units, and it was used to identify normative patterns of information flow in conversations between physicians and seriously ill patients~\cite{2020CLA}.
Although not a direct model of conversation itself, one study modeled emotional changes between a therapist and a client using two-dimensional ordinary differential equations (ODEs), and examined how differences in parameters affect emotional attractors~\cite{2011LIE}.
A classic study by Simmel (1902) examined how changes in the number of individuals within an interacting group affect the resulting social structures~\cite{1902SIM}.

\vspace{2mm}
Amid this substantial body of conversation research, the characteristics of silence have occasionally been addressed; however, few studies have undertaken an in-depth analysis of silence itself as a primary subject.

\vspace{2mm}
In this study, we propose a mathematical model of silence in co-present groups and analyze its behavior. 
Specifically, we describe the dynamics of conversation using a Markov chain model, in which each participant is represented as an agent with discrete states of silence ($0$) and speaking ($1$), incorporating mutual interactions among them. 
We then aim to reveal the properties of silence by closely examining several representative interaction patterns. 
This analysis revealed a phase-transition-like phenomenon, in which the probability of silence abruptly drops to zero once individuals reduce their awareness of surrounding conversation beyond a certain threshold. 
These findings suggest that, for silence to occur in a co-present group—namely, for an angel to pass—each individual must maintain at least a minimal level of awareness of their surroundings. 
Additionally, in order to discuss the connection between this model and the real world, we conducted a preliminary comparison between the model and experimental data, though the analysis remains limited. 

\vspace{2mm}
This study focuses on silence, which plays a significant role in human communication, and proposes a new framework for analyzing it.
This framework not only provides a deeper understanding of conversation itself, but also has the potential to contribute to intercultural communication, organizational productivity, and medical practice.

\vspace{2mm}
\section{Model and Methodology}
\label{sec.method}
In this section, we introduce a novel conversational model designed to analyze silence within co-present groups.
The model assumes a setting in which multiple agents can exist, each taking on one of two discrete states: silence ($0$) or speaking ($1$).
The next state of each agent is determined by its current state and interactions with other agents, and the resulting temporal evolution is interpreted as a form of conversation.
This can also be regarded as a natural extension of the model proposed by H. B. Kekre {\it et al.}, in which agent interactions are explicitly incorporated. 

\vspace{2mm}
\subsection{The Case of a Single Agent}
\label{subsec.isolated}
In the model proposed here, each agent is assumed to have a tendency to remain in its current state. 
That is, an agent currently in a state of silence is likely to remain silent in the next time step, whereas an agent currently speaking is likely to continue speaking. 

\vspace{2mm}
Let $s(t) \in \{0, 1\}$ denote the state of an agent at time $t$, where $s(t) = 0$ represents silence and $s(t) = 1$ represents speaking.
Let $\alpha$ and $\beta$ denote the transition probabilities from silence to speaking and from speaking to silence, respectively. We assume that $0 < \alpha, \beta \ll 1$. 

\begin{align*}
  s\left(t+1\right) = 
  \begin{cases}
  1, & \text{with probability } \left(1-\alpha-\beta\right)s\left(t\right)+\alpha \\
  0, & \text{with probability } \left(\alpha+\beta-1\right)s\left(t\right)+1-\alpha
  \end{cases}
\end{align*}

\vspace{2mm}
In general, for a model in which a state changes with a fixed probability $p$ at each time step, the expected duration for which the state persists is given by the following equation:

\begin{align}
  T = \sum_{n=1}^{\infty} n\left(1-p\right)^{n-1} p = \cfrac{1}{p} 
  \label{eq.lifetime_ave}
\end{align}

\vspace{2mm}
Therefore, in this model, the expected duration of silence and the expected duration of speaking are given by $1/\alpha$ and $1/\beta$, respectively. 
In the steady state, the probabilities of being in silence $f(s = 0)$ and speaking $f(s = 1)$ can be considered proportional to the expected durations of those states. Thus, we obtain:

\begin{align*}
  f(s) = \frac{\alpha}{\alpha+\beta} s +  \frac{\beta}{\alpha+\beta} (1-s)
\end{align*}


\subsection{The Case of $N$ Agents ($N \ge 1$)} 
We now extend the model from the previous section to the case of $N$ agents. 

\vspace{2mm}
We further assume that the next state of each agent is also influenced by its surroundings. 
In other words, when the surroundings are lively, agents are more likely to speak, whereas when the environment is quiet, they are more likely to remain silent. 

\vspace{2mm}
To formalize this, we define the following quantity $S$:

\begin{align*}
  S_i\left(t\right) = \sum_{j=1}^{N} w_{ij} s_j(t)
\end{align*}

\vspace{2mm}
Here, $s_j(t)$ denotes the state of agent $j$ at time $t$, and $w_{ij}$ represents the element of the interaction matrix that quantifies the degree of influence from agent $j$ to agent $i$.  
In addition, $w_{ij}$ is assumed to satisfy the conditions $0 \leq w_{ij}$ and $\sum_{j=1}^{N} w_{ij} = 1$. 

\vspace{2mm}
Using this definition, the state transition of agent $i$ is given by: 

\begin{align}
  s_i \left(t+1\right) = 
  \begin{cases}
  1, & \text{with probability } \left(1-\alpha-\beta\right)S_i\left(t\right)+\alpha \\
  0, & \text{with probability } \left(\alpha+\beta-1\right)S_i\left(t\right)+1-\alpha
  \end{cases}
  \label{eq.3.2.1}
\end{align}

\vspace{2mm}
If $w_{ij} = \delta_{ij}$, this corresponds to the case where each agent completely ignores its surroundings and updates its state solely based on its own current state. In this case, the model is identical to that described in the previous section. 

\vspace{2mm}
The interaction weights $w_{ij}$ characterize conversational dynamics in this model. 
For instance, even among individuals belonging to the same conversation group, the degree of mutual influence is not necessarily symmetric.  
Even within the same physical space, an imbalance of power regarding silence has been noted between students and teachers~\cite{1998JAW}.  
Moreover, a study on family conversations found that the frequency of dialogue is affected by the social distance and power relations among participants~\cite{2018WIL}. In certain situations, responses may even be deliberately withheld as a result. 
The simplest way to incorporate such imbalances is to represent $w$ as an asymmetric matrix. 

\subsection{Probability Distribution of Speaking Agents}

In this section, we analyze the probability distribution of the number of speaking agents at time $t$ in a system of $N$ agents.

\subsubsection{Isolated Group}

First, we focus on the case where $w_{ij} = \delta_{ij}$, namely, a situation in which each agent completely ignores its surroundings. 

\vspace{2mm}
In this case, the probability that $n_1$ agents are speaking after sufficient time has passed is given by the following binomial distribution: 

\begin{align*}
  f_\mathrm{iso}(n_1) = \binom{N}{n_1} \left(\frac{\alpha}{\alpha+\beta}\right)^{n_1} \left(\frac{\beta}{\alpha+\beta}\right)^{N-n_1}
\end{align*}


\subsubsection{Uniform Group}
\label{subsec.uniform}

Next, we examine the case where $w_{ij} = 1/N$.  
This corresponds to a situation in which each agent is equally influenced by all agents in the group, including itself. 
Under this assumption, the transition probabilities to the next state take the same form for all agents. 

\vspace{2mm}
Let us consider the case in which $n_0$ agents are silent and $n_1$ are speaking at time $t$, and in the next moment ($t + \Delta t$), $d_{01}$ agents begin speaking while $d_{10}$ agents become silent.  
The probability of such a transition, denoted $P(d_{01}, d_{10}; N, n_0, n_1)$, is given by:

\begin{align*}
  P(d_{01}, d_{10}; N, n_0, n_1) = 
  \binom{n_0}{d_{01}} \binom{n_1}{d_{10}} 
  p_{00}^{n_0-d_{01}}
  p_{01}^{d_{01}}
  p_{10}^{d_{10}}
  p_{11}^{n_1-d_{10}}
\end{align*}

\vspace{2mm}
Here, $p_{uv}$ represents the transition probability from state $u$ to state $v$.  
From Eq.~(\ref{eq.3.2.1}), these probabilities are expressed as:

\begin{align*}
  p_{00} &= p_{10} = \frac{n_0}{N}\left(1-\alpha\right) + \frac{n_1}{N}\beta \\
  p_{01} &= p_{11} = \frac{n_0}{N}\alpha + \frac{n_1}{N}\left(1-\beta\right) 
\end{align*}


\vspace{2mm}
The probability that the number of speaking agents changes from $n_1$ at time $t$ to $n_1'$ at time $t + \Delta t$, denoted $g(n_1, n_1')$, is then given by: 

\begin{align*}
  g(n_1, n_1') = 
  \sum_{\left(d_{01}, d_{10}\right) \in D}
  \binom{n_0}{d_{01}} \binom{n_1}{d_{10}} 
  \left( \frac{n_0}{N}\alpha + \frac{n_1}{N}\left(1-\beta\right) \right)^{n_1'}
  \left( \frac{n_0}{N}\left(1-\alpha\right) + \frac{n_1}{N}\beta \right)^{N-n_1'} &\\
  \left( {\rm where} \quad D = \left\{ \left(d_{01}, d_{10}\right) | d_{01}-d_{10}=n_1'-n_1 \right\} \right) &
\end{align*}


\vspace{2mm}
Evaluating this sum yields: 

\begin{align*}
  g(n_1, n_1') = 
  \binom{N}{n_1'}
  \left( \frac{n_0}{N}\alpha + \frac{n_1}{N}\left(1-\beta\right) \right)^{n_1'} 
  \left( \frac{n_0}{N}\left(1-\alpha\right) + \frac{n_1}{N}\beta \right)^{N-n_1'}
\end{align*}


\vspace{2mm}
This result makes use of the identity: 

\begin{align*}
  \sum_{\left(d_{01}, d_{10}\right) \in D}
  \binom{n_0}{d_{01}} \binom{n_1}{d_{10}} 
  = 
  \binom{N}{n_1'}
\end{align*}


\vspace{2mm}
As a result, the probability distribution $f_{\mathrm{uni}}(n_1')$ in the steady state satisfies the following equation: 

\begin{align*}
  f_\mathrm{uni}(n_1') = \sum_{n_1=0}^{N} g(n_1, n_1') f_\mathrm{uni}(n_1)
\end{align*}


\vspace{2mm}
In general, solving this equation analytically is difficult.  
However, by rewriting it as an integral equation and assuming that $0 < \alpha, \beta \ll 1$, we obtain the following approximate solution: 

\begin{align*}
  f_\mathrm{uni}(n_1) = 
  \frac{1}{\mathrm{Z}(2\alpha N, 2\beta N)} 
  \left\{ \left(1-\frac{n_1}{N}\right)\alpha + \frac{n_1}{N}\left(1-\beta\right) \right\}^{2\alpha N-1}  
  \left\{ \left(1-\frac{n_1}{N}\right)\left(1-\alpha\right) + \frac{n_1}{N}\beta \right\}^{2\beta N-1}
\end{align*}


\vspace{2mm}
Here, the normalization constant $\mathrm{Z}$ is defined as: 

\begin{align*}
  \mathrm{Z}(2\alpha N, 2\beta N) = 
  \sum_{n_1=0}^{N} 
  \left\{ \left(1-\frac{n_1}{N}\right)\alpha + \frac{n_1}{N}\left(1-\beta\right) \right\}^{2\alpha N-1}  
  \left\{ \left(1-\frac{n_1}{N}\right)\left(1-\alpha\right) + \frac{n_1}{N}\beta \right\}^{2\beta N-1}
\end{align*}


\vspace{2mm}
The detailed derivation of this expression is provided in the Appendix.

\vspace{2mm}
\subsubsection{Café~$\theta$ Model}

Here, we consider a formulation in which the factors determining an agent’s next state are explicitly separated into two components: the agent's own current state and the state of the surrounding environment.

\vspace{2mm}
Using a parameter $\theta$ satisfying $0 \leq \theta \leq 1$, we define $w_{ij}$ as follows: 

\begin{align}
w_{ij} = \left\{
\begin{array}{ll}
\theta & (i=j)\\
\frac{1-\theta}{N-1} & (i \ne j)
\end{array}
\right.
\label{eq.uni_iso_w}
\end{align}

\vspace{2mm}
This parameter $\theta$ represents the degree to which an agent's next state is influenced by its own current state. When $\theta = 0$, the next state is determined entirely by the environment; when $\theta = 1$, it depends solely on the agent's own state. 

\vspace{2mm}
Under this formulation of $w_{ij}$, we consider the probability $f_{\theta}(n_1)$ that $n_1$ agents are speaking after a sufficiently long time has elapsed. 
We assume that $N$ is large. 

\vspace{2mm}
When $\theta = 0$, $w_{ij}$ can be approximated as $1/N$. 
Thus, from the result in Section~\ref{subsec.uniform}, $f_{\theta=0}(n_1)$ is given by: 

\begin{align*}
  f_\mathrm{\theta=0}(n_1) = 
  \frac{1}{\mathrm{Z}(2\alpha N, 2\beta N)} 
  \left\{ \left(1-\frac{n_1}{N}\right)\alpha + \frac{n_1}{N}\left(1-\beta\right) \right\}^{2\alpha N-1}  
  \left\{ \left(1-\frac{n_1}{N}\right)\left(1-\alpha\right) + \frac{n_1}{N}\beta \right\}^{2\beta N-1}
\end{align*}


\vspace{2mm}
In contrast, when $\theta = 1$, we have $w_{ij} = \delta_{ij}$, which corresponds exactly to the isolated case in Section~\ref{subsec.isolated}. In this case, the distribution becomes: 

\begin{align*}
  f_\mathrm{\theta=1}(n_1) 
  = \binom{N}{n_1} \left(\frac{\alpha}{\alpha+\beta}\right)^{n_1} \left(\frac{\beta}{\alpha+\beta}\right)^{N-n_1}
\end{align*}


\vspace{2mm}
Given that $N$ is large and $1 \ll \alpha N / (\alpha + \beta)$, this binomial distribution can be approximated by a normal distribution: 

\begin{align*}
  f_\mathrm{\theta=1}(n_1) 
  &= \binom{N}{n_1} \left(\frac{\alpha}{\alpha+\beta}\right)^{n_1} \left(\frac{\beta}{\alpha+\beta}\right)^{N-n_1}\\
  &= \binom{N}{n_1} \left(\frac{\alpha}{\alpha+\beta}\right)^{n_1} \left(1-\frac{\alpha}{\alpha+\beta}\right)^{N-n_1}\\
  &\sim 
  \frac{1}{\sqrt{2\pi \frac{\alpha\beta N}{(\alpha+\beta)^2}}}
  \exp \left( -\frac{\left(n_1-\frac{\alpha N}{\alpha+\beta}\right)^2}{2\frac{\alpha\beta N}{(\alpha+\beta)^2}} \right)
\end{align*}

\vspace{2mm}
Introducing the scaled variable $n_1/N$, we obtain an approximation by a Beta distribution: 

\begin{align*}
  f_\mathrm{\theta=1}(n_1) 
  &\sim 
  \frac{1}{\sqrt{2\pi \frac{\alpha\beta}{(\alpha+\beta)^2}\frac{1}{N}}}
  \exp \left( -\frac{\left(\frac{n_1}{N}-\frac{\alpha}{\alpha+\beta}\right)^2}{2\frac{\alpha\beta}{(\alpha+\beta)^2}\frac{1}{N}} \right)\\
  &\sim 
  \mathrm{B}\left(\frac{\alpha N}{\alpha+\beta}, \frac{\beta N}{\alpha+\beta}\right)
\end{align*}

\vspace{2mm}
Furthermore, under the assumption $0 < \alpha, \beta \ll 1$, we have the following approximations: 

\begin{align*}
  \frac{n_1}{N} &\sim \left(1-\frac{n_1}{N}\right)\alpha + \frac{n_1}{N}\left(1-\beta\right)\\
  1-\frac{n_1}{N} &\sim \left(1-\frac{n_1}{N}\right)\left(1-\alpha\right) + \frac{n_1}{N}\beta
\end{align*}

\vspace{2mm}
Substituting these into the previous expression, we obtain: 

\begin{align*}
  f_\mathrm{\theta=1}(n_1) \sim 
  \frac{1}{Z\left(\frac{\alpha N}{\alpha+\beta}, \frac{\beta N}{\alpha+\beta}\right)} 
  \left\{ \left(1-\frac{n_1}{N}\right)\alpha + \frac{n_1}{N}\left(1-\beta\right) \right\}^{\frac{\alpha N}{\alpha+\beta}-1}  
  \left\{ \left(1-\frac{n_1}{N}\right)\left(1-\alpha\right) + \frac{n_1}{N}\beta \right\}^{\frac{\beta N}{\alpha+\beta}-1}
\end{align*}


\vspace{2mm}
Based on these results, we assume the general form of $f_{\theta}(n_1)$ to be: 

\begin{align}
  f_{\theta}(n_1) = 
  \frac{1}{Z\left(h_\alpha(\theta), h_\beta(\theta)\right)} 
  \left\{ \left(1-\frac{n_1}{N}\right)\alpha + \frac{n_1}{N}\left(1-\beta\right) \right\}^{h_\alpha(\theta)-1}  
  \left\{ \left(1-\frac{n_1}{N}\right)\left(1-\alpha\right) + \frac{n_1}{N}\beta \right\}^{h_\beta(\theta)-1}
  \label{eq.f_theta}
\end{align}

\vspace{2mm}
Here, $h_\alpha(\theta)$ and $h_\beta(\theta)$ are functions that satisfy the following boundary conditions: 

\begin{table}[h]
  \centering
  \begin{tabular}{c|ccc}
  $\theta$ & $0$ & $\to$ & $1$ \\
  \hline
  $h_\alpha(\theta)$ & $2\alpha N$ & & $\frac{\alpha N}{\alpha + \beta}$ \\
  $h_\beta(\theta)$  & $2\beta N$  & & $\frac{\beta N}{\alpha + \beta}$  \\
  \end{tabular}
  \caption{Boundary conditions for $h_\alpha(\theta)$ and $h_\beta(\theta)$.}
\end{table}

\vspace{2mm}
Equation~(\ref{eq.f_theta}) is derived under the assumption that $N$ is sufficiently large. 
Therefore, this distribution is not intended for modeling one-on-one conversations, but rather applies to situations in which many individuals coexist in the same physical space—such as in cafes, restaurants, libraries, or classrooms. 
Accordingly, we refer to this formulation as the {\it Café~$\theta$ Model} in the present paper. 

\vspace{2mm}
\section{Results and Discussion}
\label{sec.results}

In the following sections, we present our analysis based on the Café~$\theta$ model. 

\subsection{Interpolating Function $h$}

We begin by determining the interpolating functions $h_\alpha(\theta)$ and $h_\beta(\theta)$ such that they best match the results obtained from simulations. 

\vspace{2mm}
To begin, we assume the forms of the interpolating functions as follows, in order to satisfy the boundary conditions: 

\begin{align}
  h_\alpha(\theta) &= N\alpha\left(2\left(1-h(\theta)\right)+\frac{h(\theta)}{\alpha+\beta}\right)\nonumber\\
  h_\beta(\theta)  &= N\beta\left(2\left(1-h(\theta)\right)+\frac{h(\theta)}{\alpha+\beta}\right)
  \label{eq.h_approx}
\end{align}

\vspace{2mm}
Here, $h(\theta)$ is a function that satisfies $h(0) = 0$ and $h(1) = 1$. While its explicit form is omitted for simplicity, it in fact depends on the parameters $(N, \alpha, \beta)$.
To examine this dependency, we conducted simulations across various parameter combinations and evaluated the most plausible value of $h$ for each setting. 
Specifically, for each parameter set, we computed the Kullback–Leibler (KL) divergence between the distribution obtained from the simulation and that given by Eq.~(\ref{eq.f_theta}). The value of $h$ that minimizes this divergence is denoted as $h_{\mathrm{min}}$ and treated as the most likely candidate.

\vspace{2mm}
In this study, we used the parameter settings listed in Table~\ref{tab.sim_param}. 

\begin{table}[htbp]
  \centering
    \caption{Parameter settings used in the simulations}
  \label{tab.sim_param}
  \begin{tabular}{ll}
  \hline
  \textbf{Parameter} & \textbf{Range} \\
  \hline
  $N$ & \( 512, 1024, 2048, 4096 \) \\
  $\alpha$ & \( \frac{1}{100}, \frac{1}{200}, \frac{1}{400}, \frac{1}{800}, \frac{1}{1600}, \frac{1}{3200}, \frac{1}{6400} \) \\
  $\beta$ & \( \frac{1}{100}, \frac{1}{200}, \frac{1}{400}, \frac{1}{800}, \frac{1}{1600}, \frac{1}{3200}, \frac{1}{6400} \) \\
  $\theta$ & \( 0.0, 0.25, 0.50, 0.75, 0.80, 0.85, 0.90, 0.95, 1.0 \) \\
  \hline
  \end{tabular}
\end{table}

\begin{figure}[htbp]
  \centering
  \begin{subfigure}{0.49\textwidth}
    \centering
    \includegraphics[width=\linewidth]{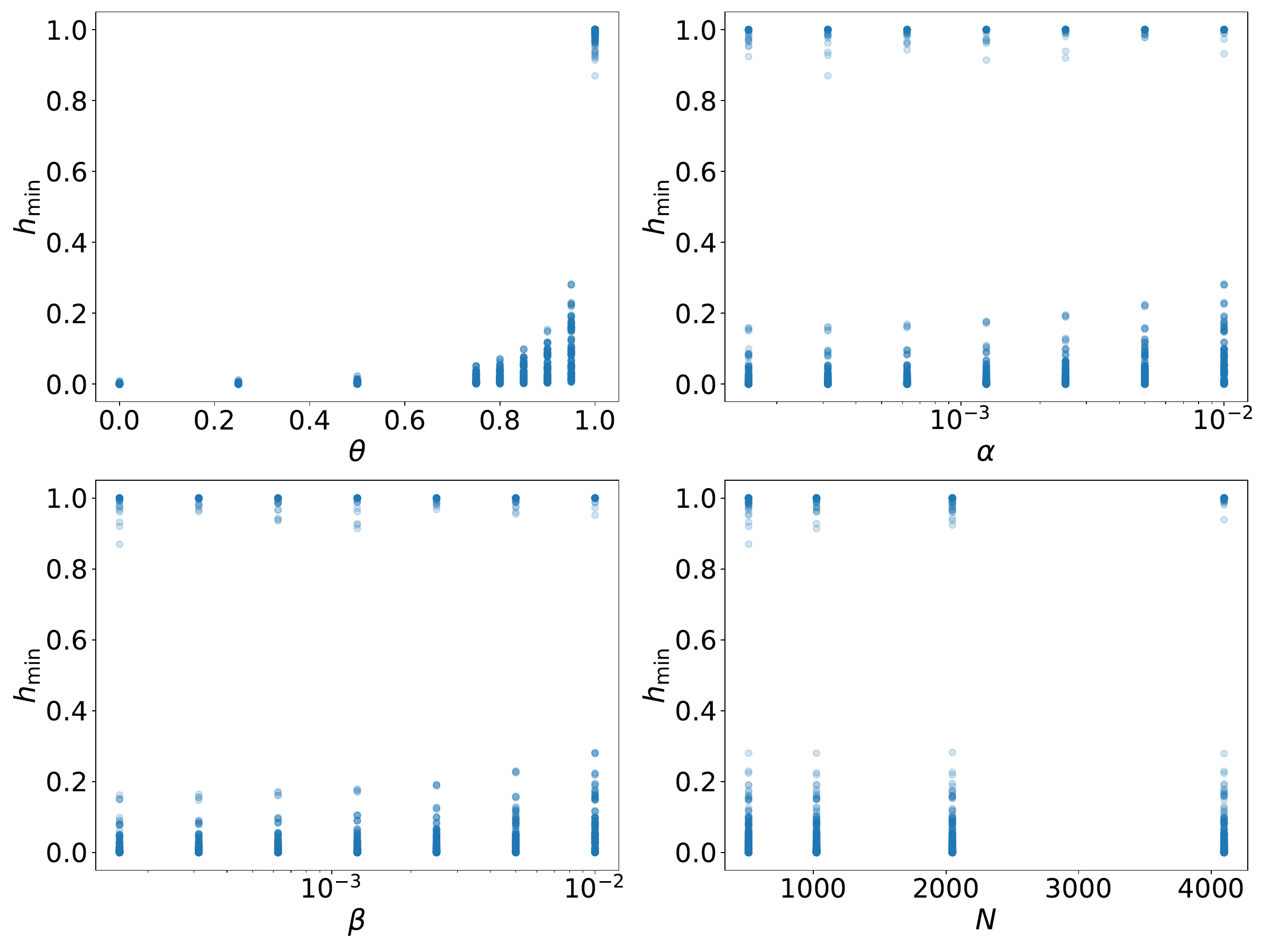}
    \caption{
      Relationship between each parameter ($N, \alpha, \beta, \theta$) and $h_{\mathrm{min}}$. 
      A positive correlation is observed only between $h_{\mathrm{min}}$ and $\theta$. 
      }
    \label{fig.01_a}
  \end{subfigure}
  \hfill
  \begin{subfigure}{0.49\textwidth}
    \centering
    \includegraphics[width=\linewidth]{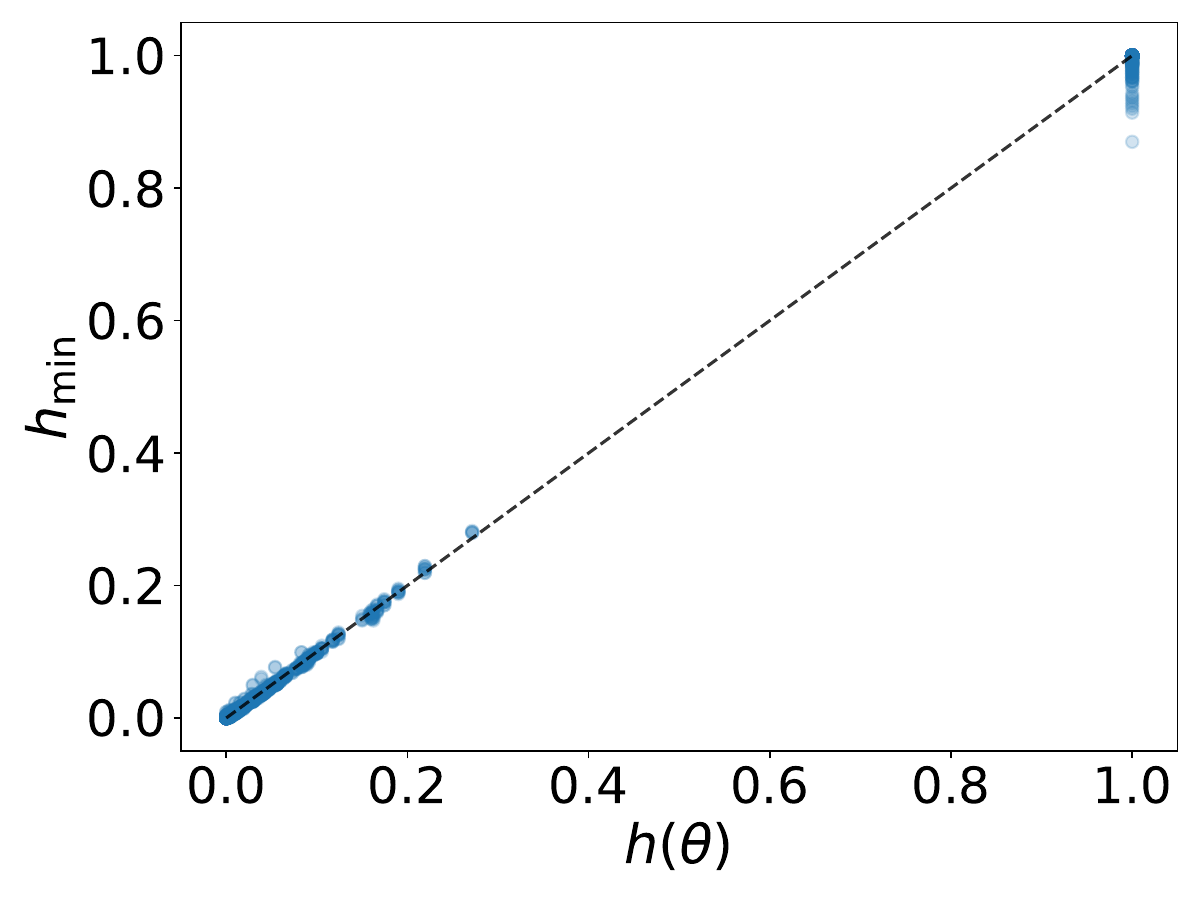}
    \caption{
      Comparison between the regression model obtained via symbolic regression and the simulated values of $h_{\mathrm{min}}$. 
      The black dashed line represents $y = x$, showing that the model fits the simulation results well.
    }
    \label{fig.01_b}
  \end{subfigure}
  \caption{
    Relationship between the simulation parameters, the regression model, and $h_{\mathrm{min}}$. 
  }
  \label{fig.01}
\end{figure}

\vspace{2mm}
Figure~\ref{fig.01_a} shows how $h_{\mathrm{min}}$ varies with the parameters. 
Apart from a clear positive correlation between $h_{\mathrm{min}}$ and $\theta$, no strong dependency is observed with the other parameters. 

\vspace{2mm}
Next, we applied symbolic regression using PySR~\cite{2023CRA} to the pairs of simulation parameters and corresponding values of $h_{\mathrm{min}}$. 
Among the resulting candidate functions, we selected a simple expression that satisfies the boundary conditions: 

\begin{align}
  h(\theta) \sim \frac{\theta(\alpha+\beta)}{1-\theta+\alpha+\beta}
  \label{eq.h_theta_approx}
\end{align}

\vspace{2mm}
Figure~\ref{fig.01_b} compares this expression with the simulation results, demonstrating excellent agreement. 

\vspace{2mm}
Based on these findings, we adopt the assumed form of Eq.~(\ref{eq.h_approx}) and the approximation in Eq.~(\ref{eq.h_theta_approx}) as the interpolating functions $h_\alpha(\theta)$ and $h_\beta(\theta)$ in the remainder of this paper. 

\subsection{Evaluation of Approximations}

To assess the validity of the approximations, we compare the outcomes of the Café~$\theta$ model with numerical simulation results. 

\begin{figure}[htbp]
  \centering
  \begin{subfigure}{0.32\textwidth}
    \centering
    \includegraphics[width=\linewidth]{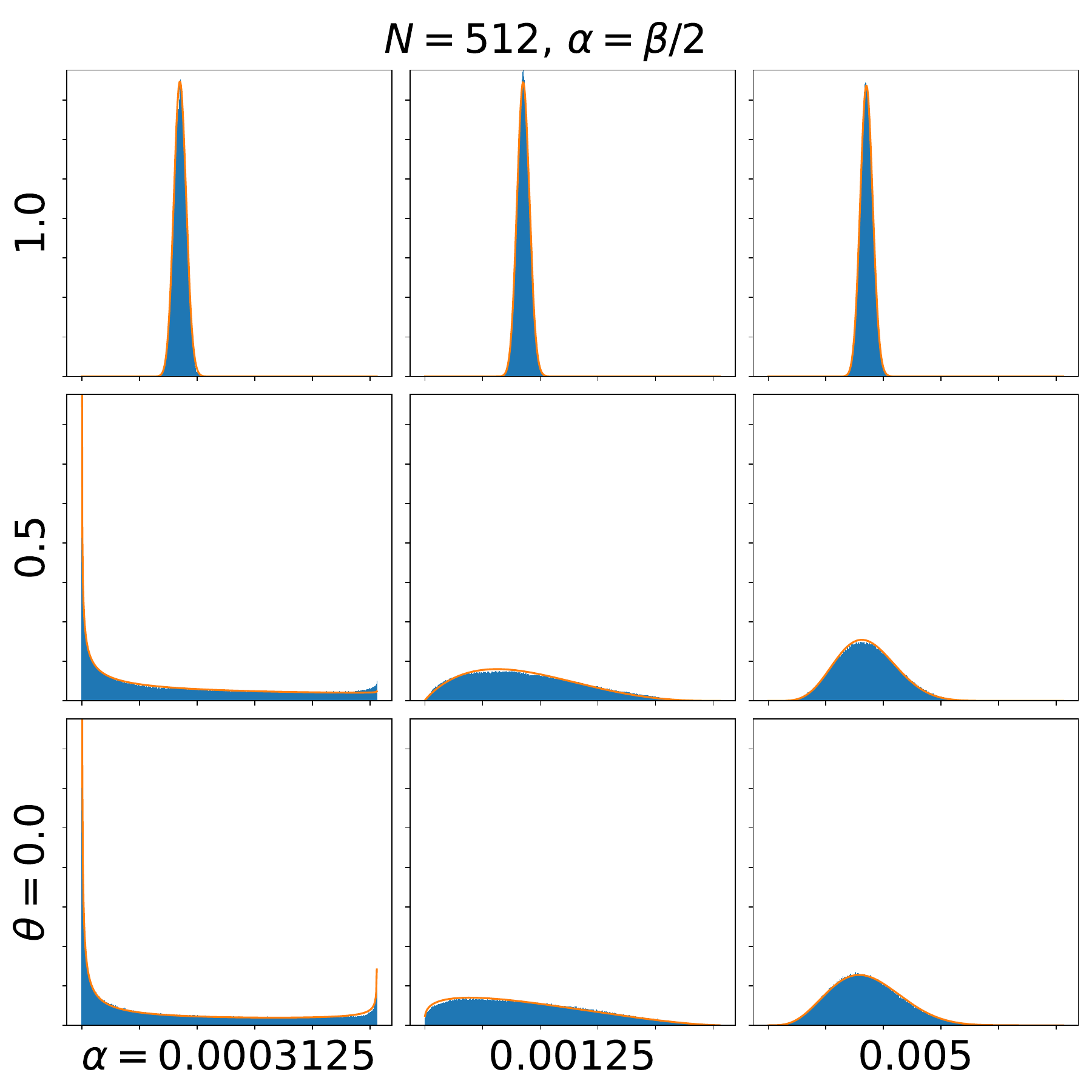}
    \caption{
      $\alpha=\beta/2$
    }
    \label{fig.02_a}
  \end{subfigure}
  \hfill
  \begin{subfigure}{0.32\textwidth}
    \centering
    \includegraphics[width=\linewidth]{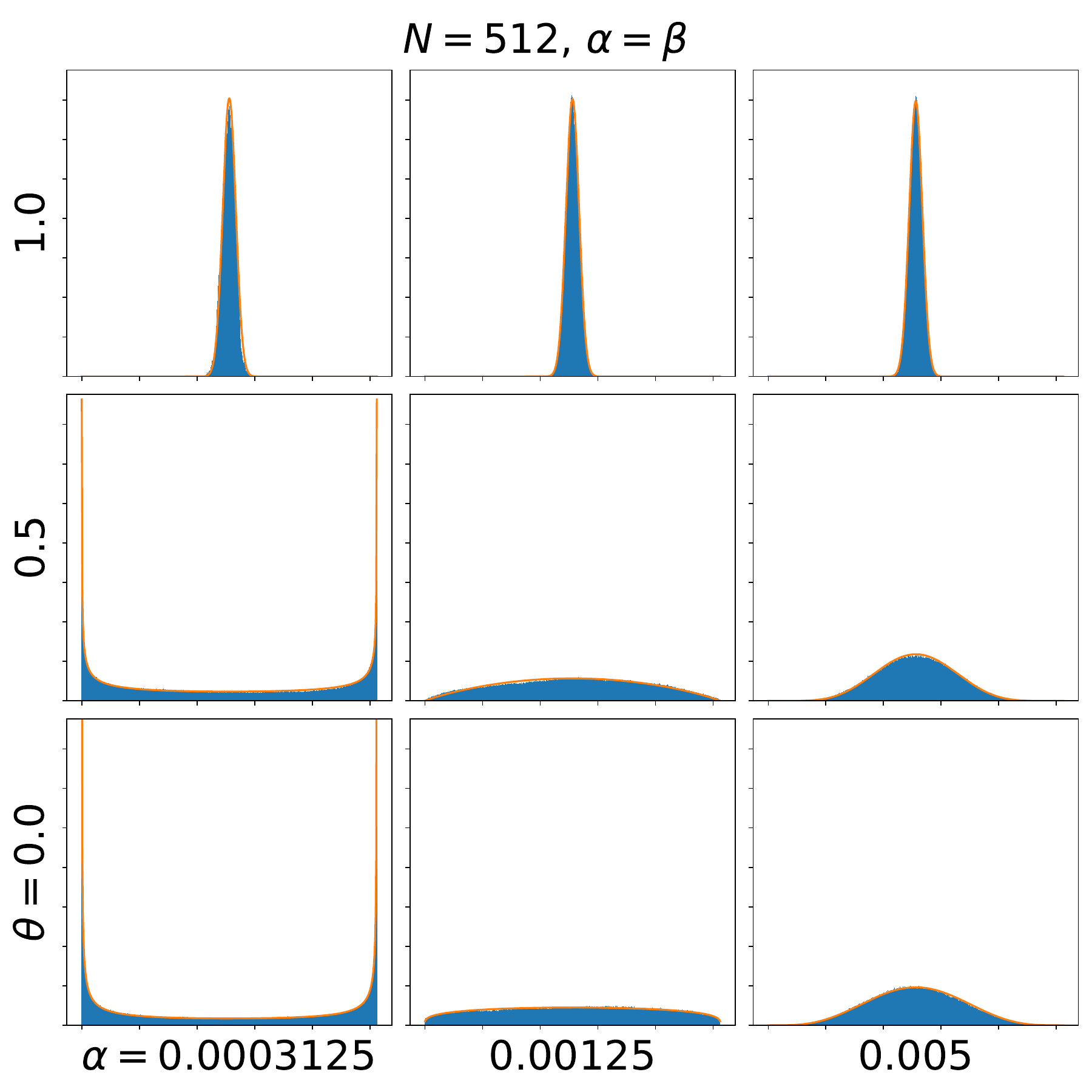}
    \caption{
      $\alpha=\beta$
    }
    \label{fig.02_b}
  \end{subfigure}
  \hfill
  \begin{subfigure}{0.32\textwidth}
    \centering
    \includegraphics[width=\linewidth]{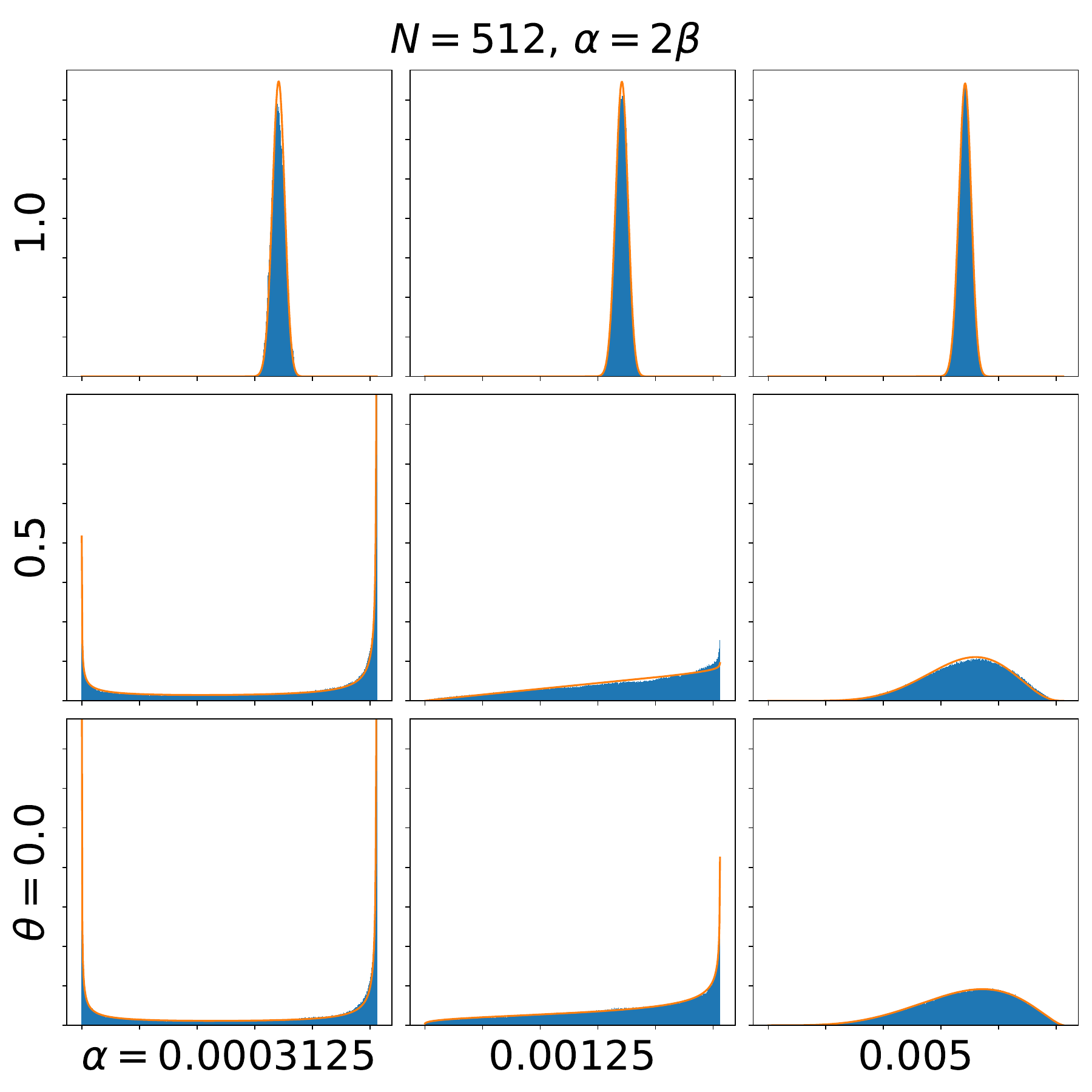}
    \caption{
      $\alpha=2\beta$
    }
    \label{fig.02_c}
  \end{subfigure}
  \caption{
    Comparison between the Café~$\theta$ model (orange line) and simulations (blue histogram). 
    Three different relationships between $\alpha$ and $\beta$ are considered, and for each case, distributions are shown for varying values of $\theta = 0.0, 0.5, 1.0$ and $\alpha = 1/3200, 1/800, 1/200$. 
    Panels (a), (b), and (c) correspond to $\alpha = \beta/2$, $\alpha = \beta$, and $\alpha = 2\beta$, respectively. 
    In all cases, the model and simulation results are in excellent agreement. 
  }
  \label{fig.02}
\end{figure}

\vspace{2mm}
We first verify the consistency between the model and simulation results for a sufficiently large number of agents. Figure~\ref{fig.02} shows the results for $N = 512$. 
For all parameter settings, the Café~$\theta$ model exhibits excellent agreement with the simulation results. 

\begin{figure}[htbp]
  \centering
  \begin{subfigure}{0.32\textwidth}
    \centering
    \includegraphics[width=\linewidth]{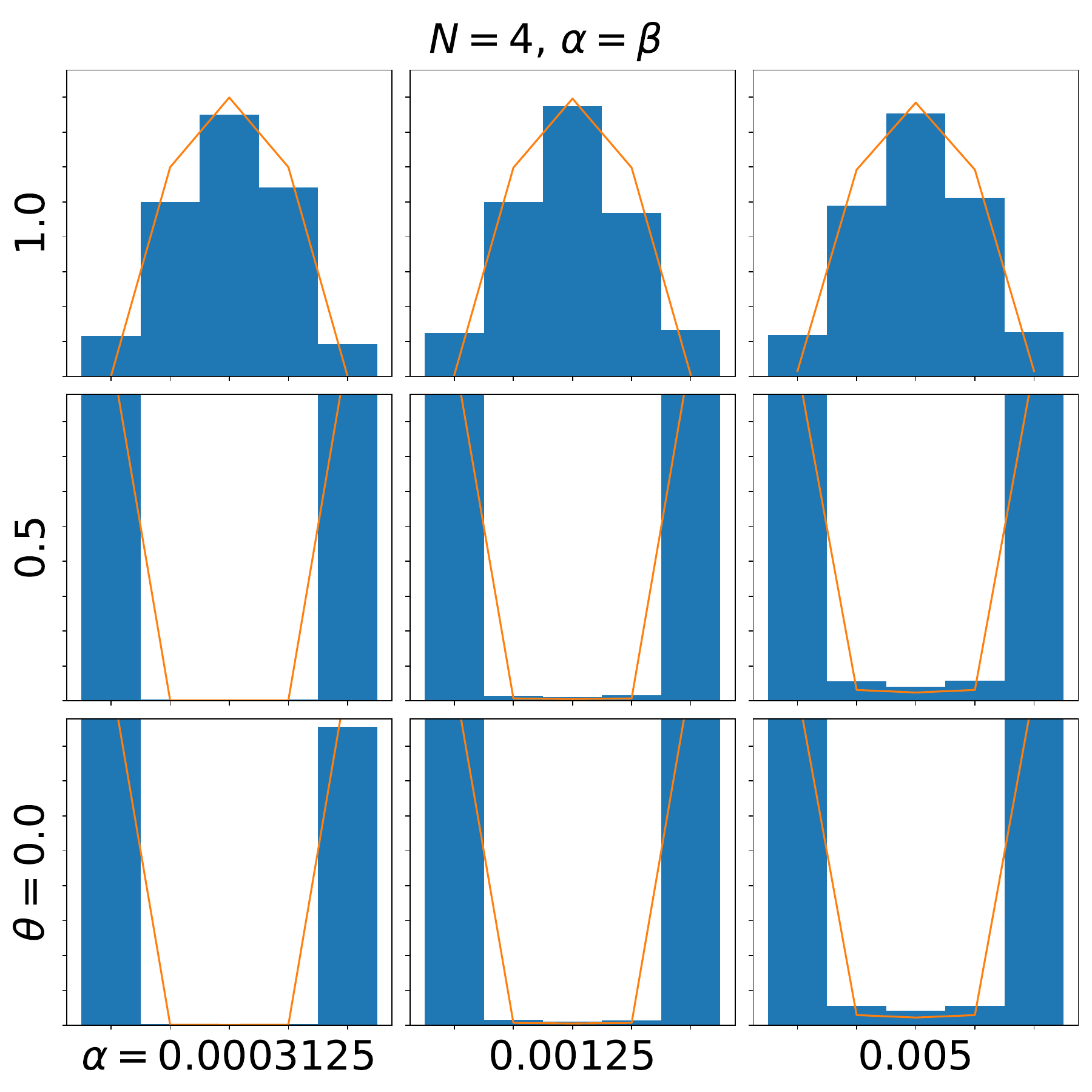}
    \caption{
      $N=4$
    }
    \label{fig.03_a}
  \end{subfigure}
  \hfill
  \begin{subfigure}{0.32\textwidth}
    \centering
    \includegraphics[width=\linewidth]{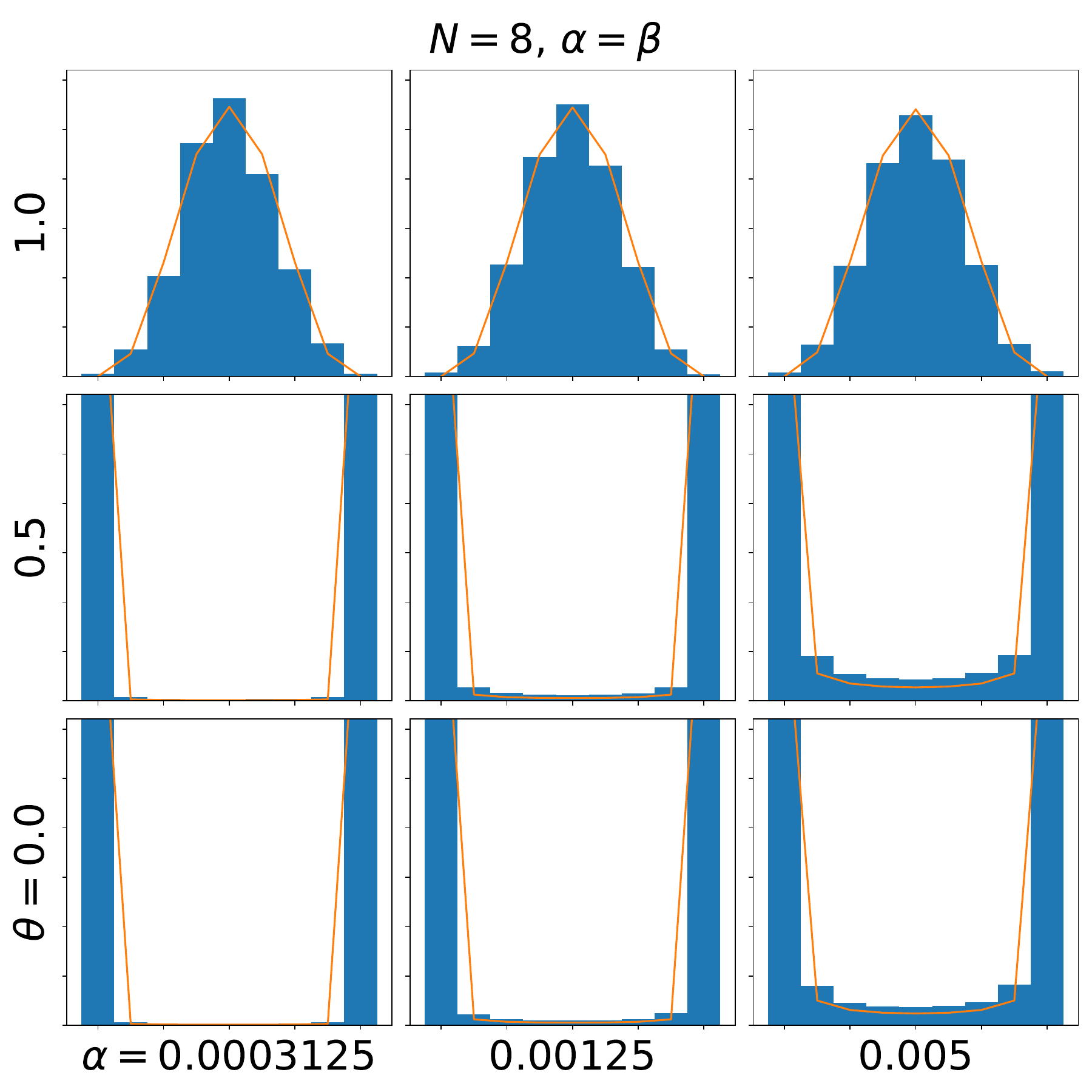}
    \caption{
      $N=8$
    }
    \label{fig.03_b}
  \end{subfigure}
  \hfill
  \begin{subfigure}{0.32\textwidth}
    \centering
    \includegraphics[width=\linewidth]{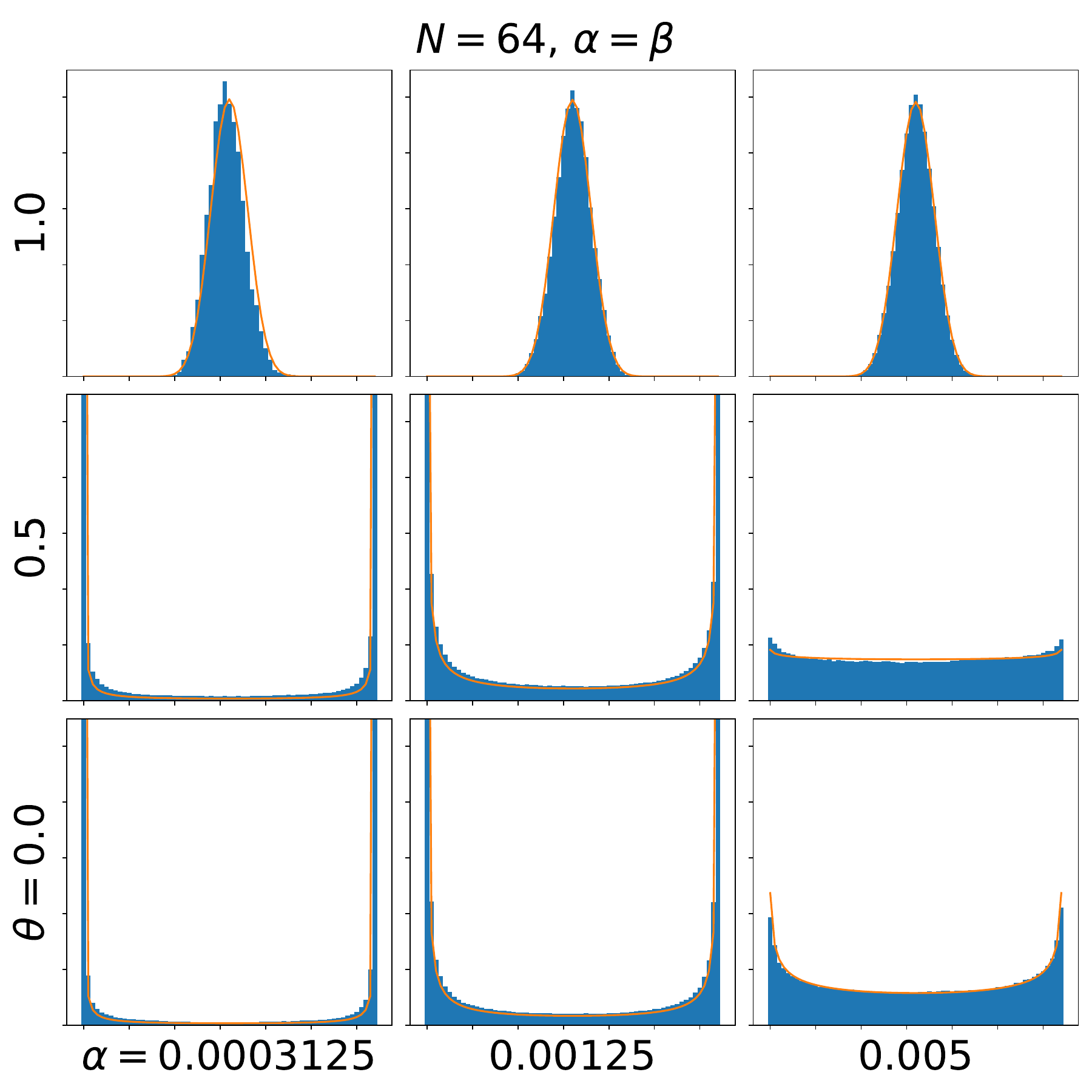}
    \caption{
      $N=64$
    }
    \label{fig.03_c}
  \end{subfigure}
  \caption{
    Comparison between the Café~$\theta$ model (orange line) and simulations (blue histogram). 
    Here we consider only the case $\alpha = \beta$, and show distributions for varying values of $\theta = 0.0, 0.5, 1.0$ and $\alpha = 1/3200, 1/800, 1/200$.
    Panels (a), (b), and (c) correspond to $N = 4$, $N = 8$, and $N = 64$, respectively. 
    The results show excellent agreement across all settings, even for small $N$. 
  }
  \label{fig.03}
\end{figure}

\vspace{2mm}
Next, we investigate how small $N$ can be while still maintaining agreement between the model and simulations. Figure~\ref{fig.03} shows the results for $N = 4, 8, 64$. 
Although the derivation of the Café~$\theta$ model assumes a large $N$, we observe that the approximation holds well even for very small values such as $N = 4$. 

\subsection{Random Weight $w$ and the Café~$\theta$ Model}

We next investigate how the Café~$\theta$ model deviates from simulation results when the interaction weight matrix $w$ is randomized. 

\vspace{2mm}
The off-diagonal elements of $w$ represent the degree of influence an agent receives from others. 
It is therefore reasonable to assume that $w_{ij}$ decreases as the distance between agents $i$ and $j$ increases. 
In this experiment, we randomly place $N$ agents on a two-dimensional plane and define $w'_{ij}$ to be inversely proportional to the square of the distance between them (with diagonal elements set to zero). 
Using this $w'_{ij}$, the interaction weights $w_{ij}$ are defined as: 

\begin{align*}
  w_{ij} = \left\{
  \begin{array}{ll}
  \theta & (i=j)\\
  (1-\theta)\frac{w'_{ij}}{\sum_k w'_{ik}} & (i \ne j)
  \end{array}
  \right.
\end{align*}

\vspace{2mm}
The positions of the $N$ agents are determined by sampling from a uniform distribution over $[0, 1)$ in both spatial dimensions. 
For each parameter setting, we generated 10 random configurations. 

\begin{figure}[htbp]
  \centering
  \begin{subfigure}{0.32\textwidth}
    \centering
    \includegraphics[width=\linewidth]{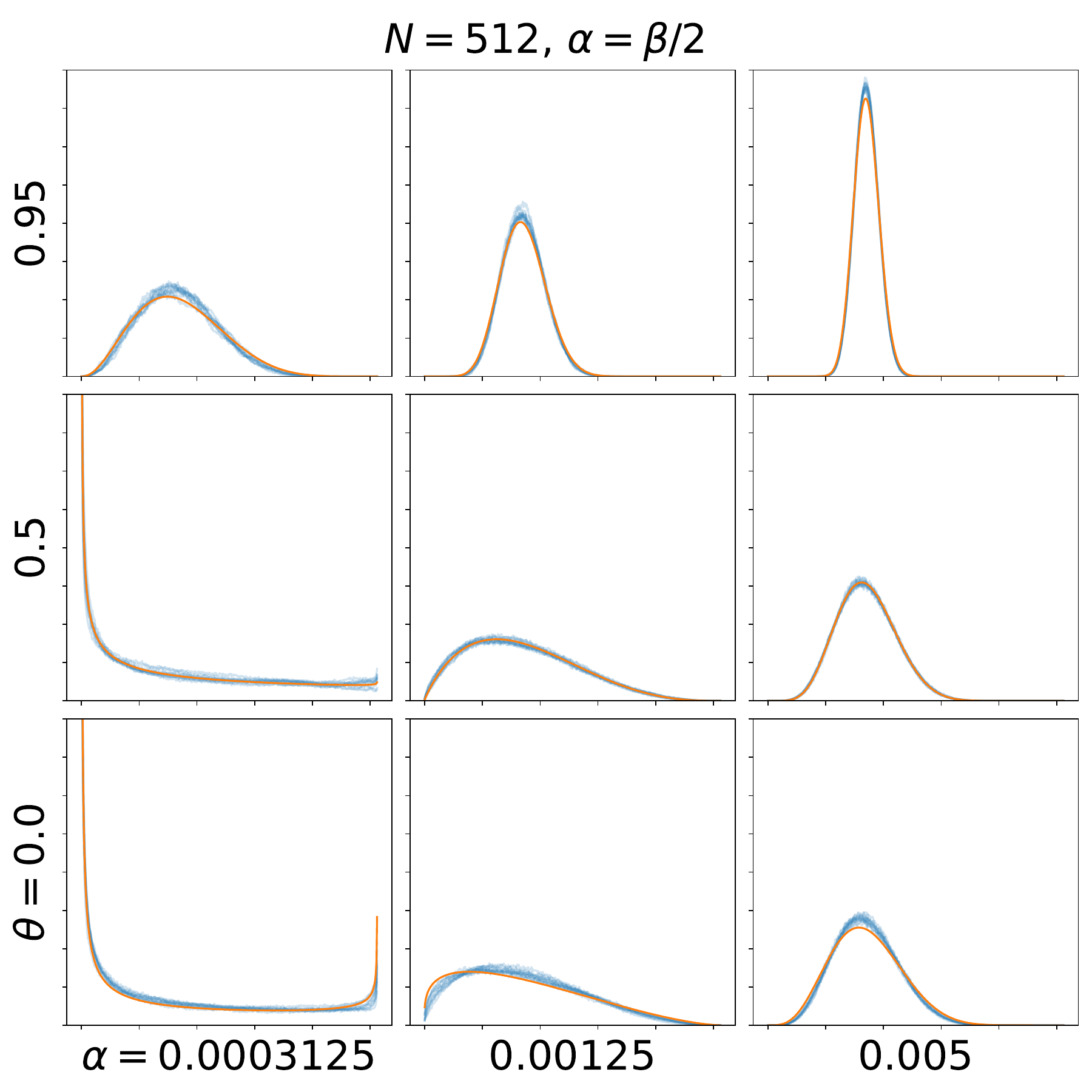}
    \caption{$\alpha=\beta/2$}
    \label{fig.04_a}
  \end{subfigure}
  \hfill
  \begin{subfigure}{0.32\textwidth}
    \centering
    \includegraphics[width=\linewidth]{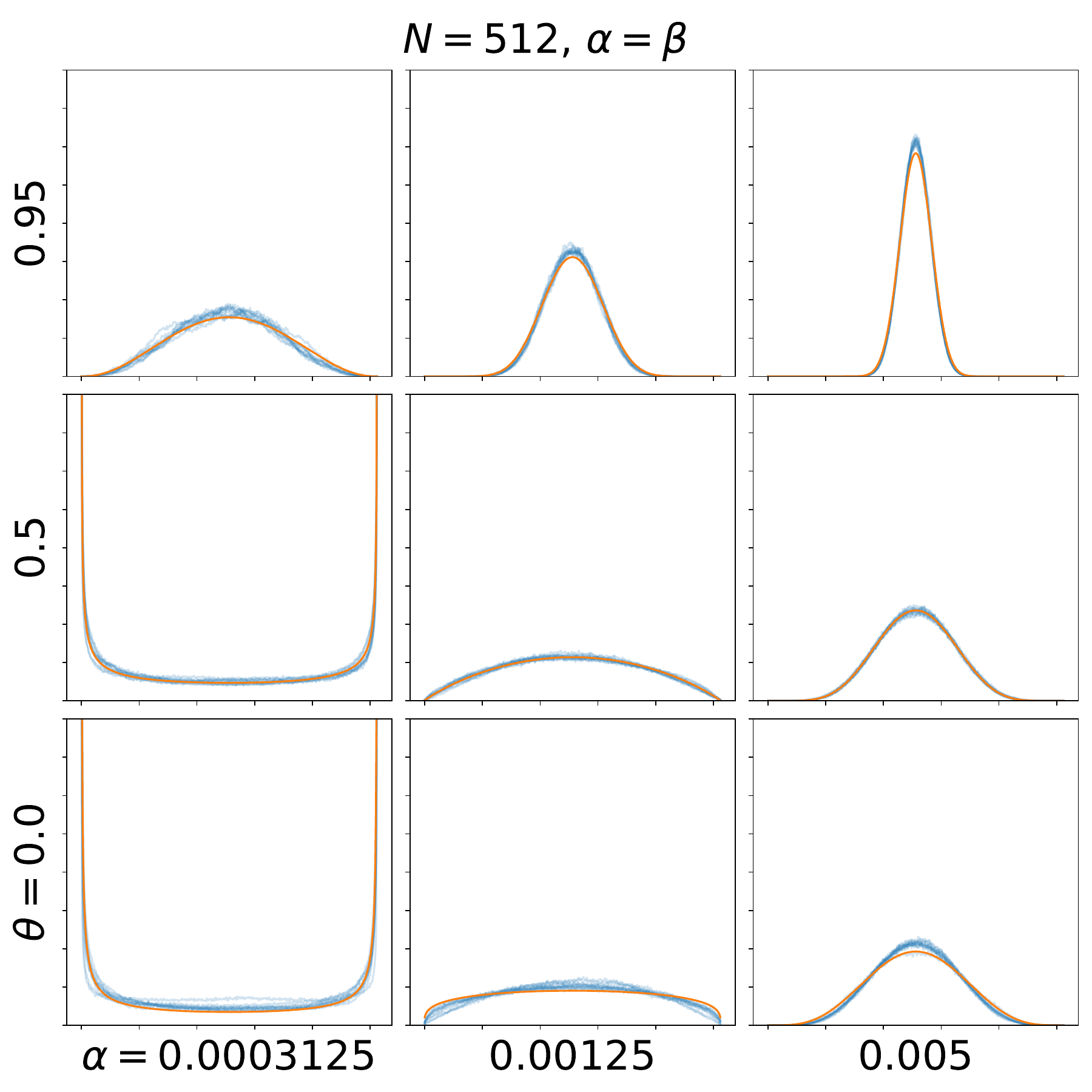}
    \caption{$\alpha=\beta$}
    \label{fig.04_b}
  \end{subfigure}
  \hfill
  \begin{subfigure}{0.32\textwidth}
    \centering
    \includegraphics[width=\linewidth]{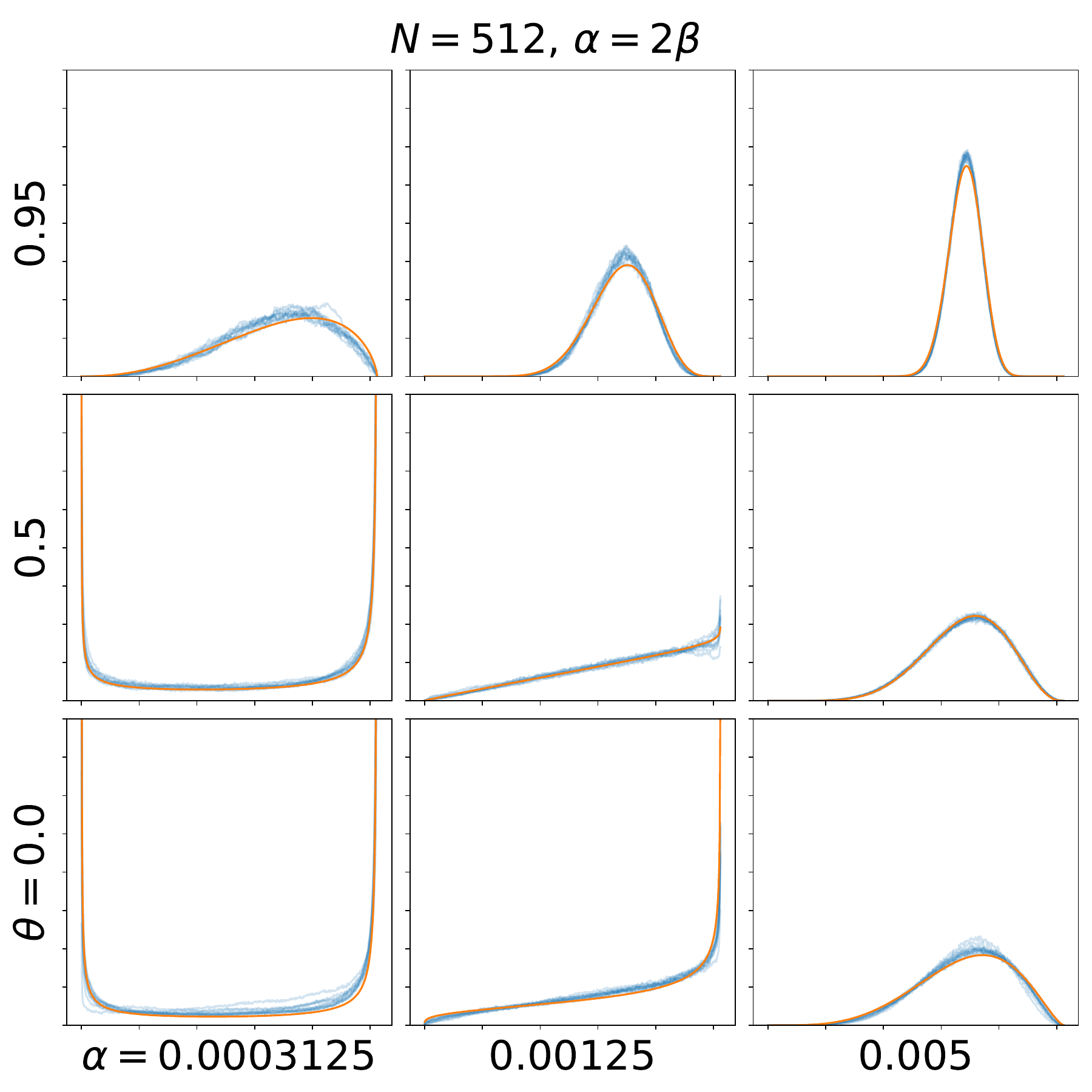}
    \caption{$\alpha=2\beta$}
    \label{fig.04_c}
  \end{subfigure}
  \caption{
    Comparison between the Café~$\theta$ model (orange lines) and simulation results (blue lines). 
    Each simulation plot shows the results of 10 random trials superimposed for comparison. 
    Three different relationships between $\alpha$ and $\beta$ are examined: (a) $\alpha = \beta/2$, (b) $\alpha = \beta$, and (c) $\alpha = 2\beta$. 
    For each case, distributions are shown under varying values of $\theta = 0.0, 0.5, 0.95$ and $\alpha = 1/3200, 1/800, 1/200$. 
    Across all parameter settings, the model demonstrates strong agreement with the simulation data. 
  }
  \label{fig.04}
\end{figure}

\vspace{2mm}
As an example of a large-scale system, the results for $N = 512$ are shown in Fig.~\ref{fig.04}. 
Despite the weights being randomly generated, the agreement between the Café~$\theta$ model and the simulation is remarkably good. 

\begin{figure}[htbp]
  \centering
  \begin{subfigure}{0.32\textwidth}
    \centering
    \includegraphics[width=\linewidth]{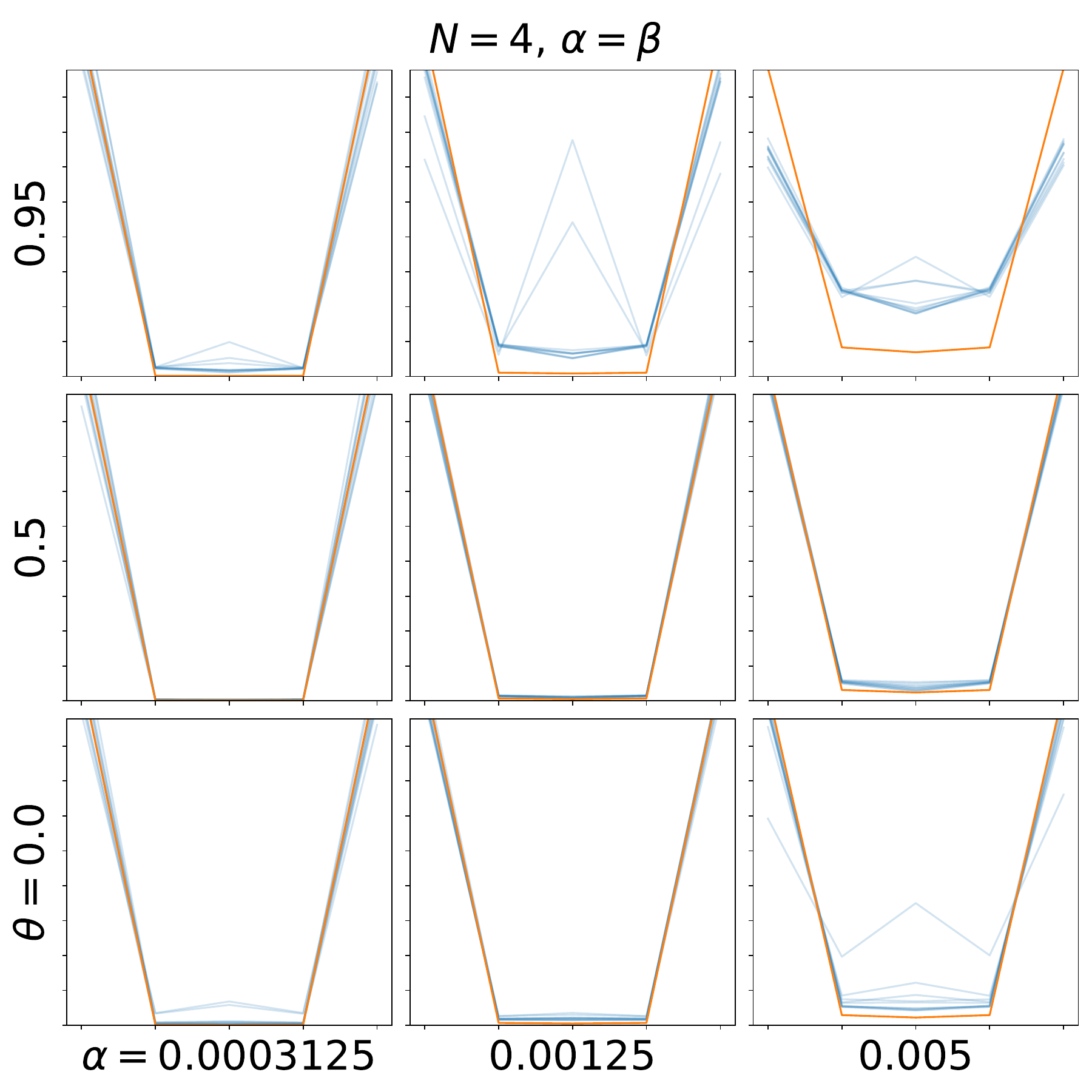}
    \caption{$N=4$}
    \label{fig.05_a}
  \end{subfigure}
  \hfill
  \begin{subfigure}{0.32\textwidth}
    \centering
    \includegraphics[width=\linewidth]{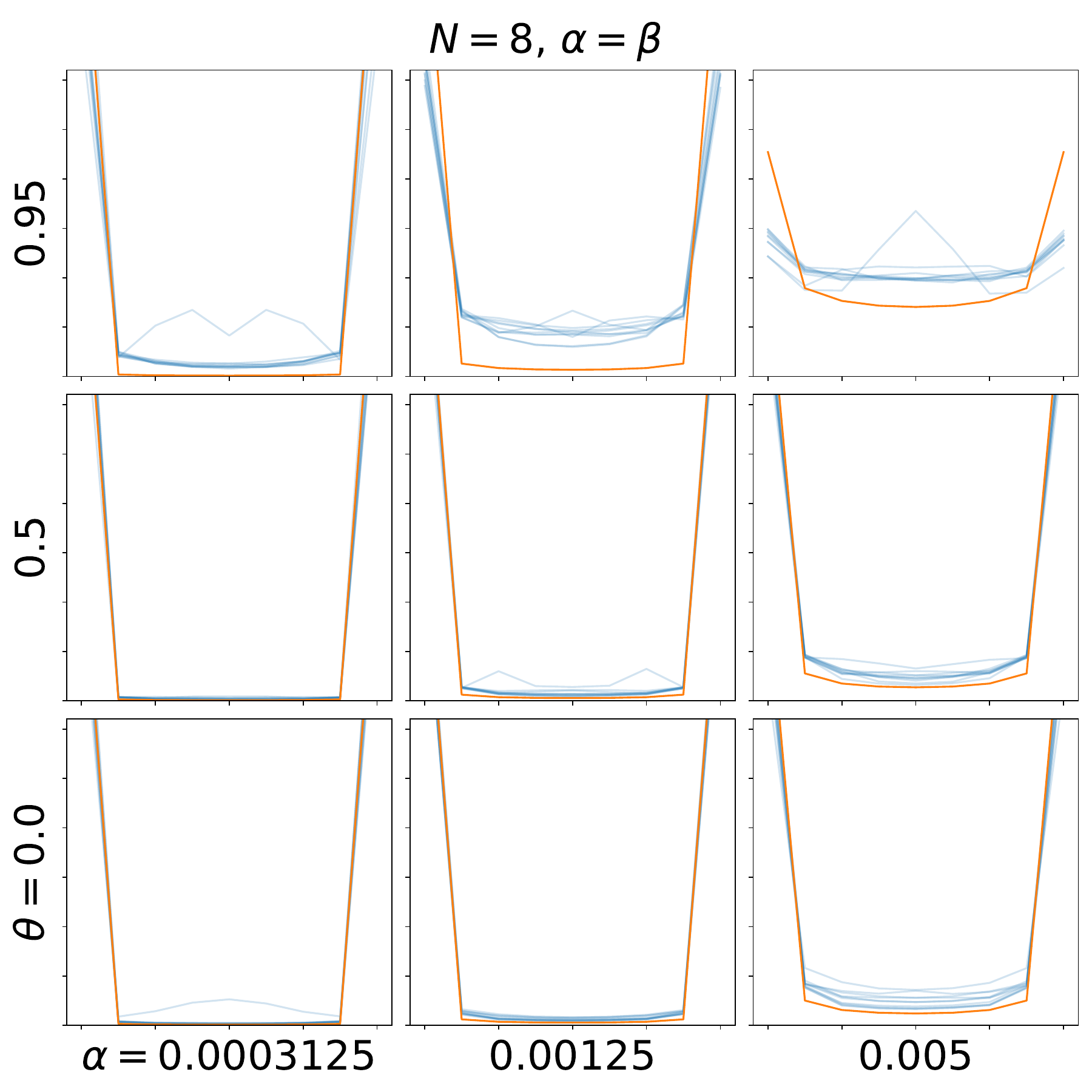}
    \caption{$N=8$}
    \label{fig.05_b}
  \end{subfigure}
  \hfill
  \begin{subfigure}{0.32\textwidth}
    \centering
    \includegraphics[width=\linewidth]{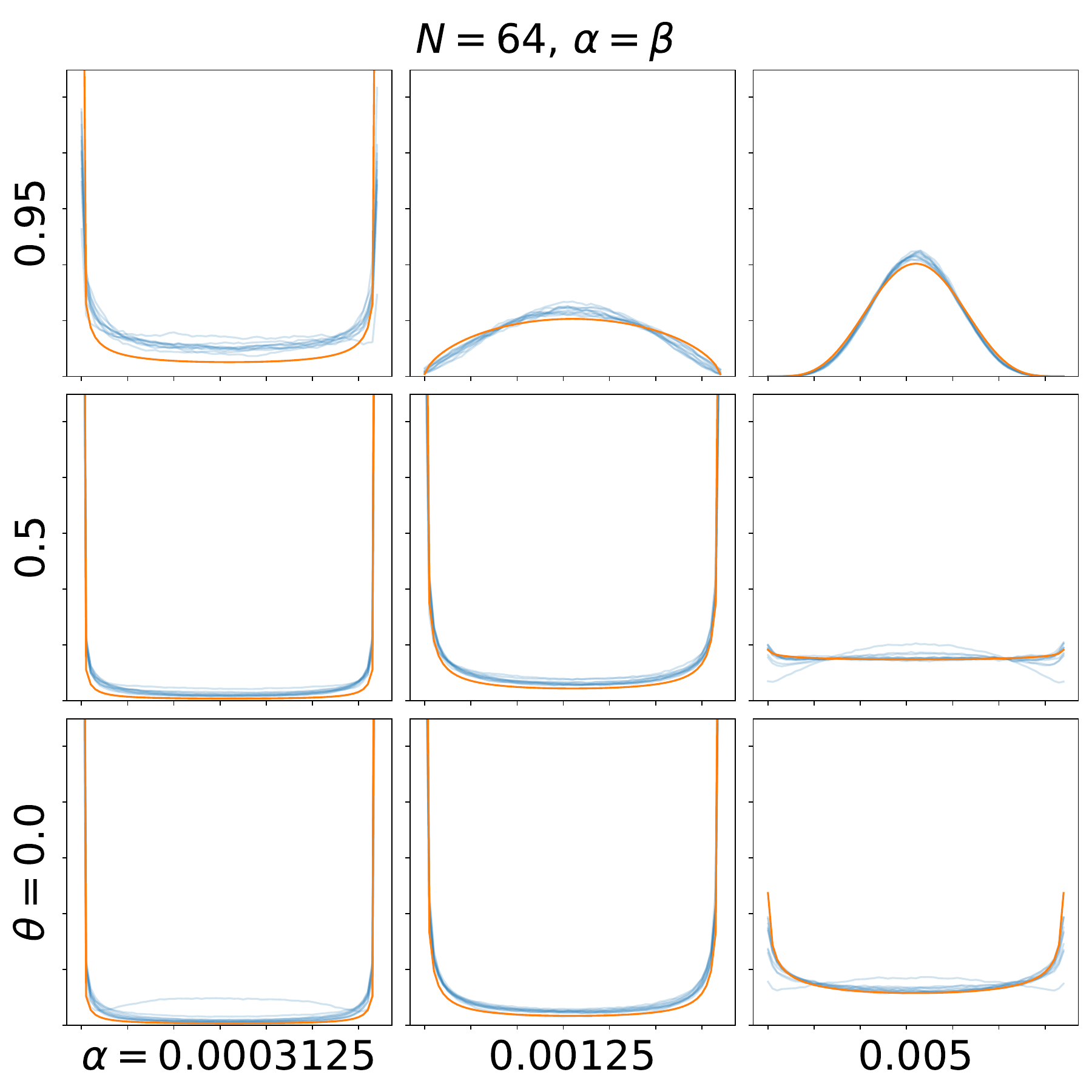}
    \caption{$N=64$}
    \label{fig.05_c}
  \end{subfigure}
  \caption{
    Comparison between the Café~$\theta$ model (orange lines) and simulation results (blue lines). 
    Each simulation plot shows 10 randomized configurations. 
    Only the case $\alpha = \beta$ is considered here, with $\theta = 0.0, 0.5, 0.95$ and $\alpha = 1/3200, 1/800, 1/200$. 
    Panels (a), (b), and (c) correspond to $N = 4$, $8$, and $64$, respectively. 
    While the model captures the overall distribution trend well, it fails to reproduce some complex shapes such as central-edge-peaked or multimodal distributions. 
  }
  \label{fig.05}
\end{figure}

\vspace{2mm}
Next, to assess performance in smaller systems, we present the results for $N = 4$, $8$, and $64$ in Fig.~\ref{fig.05}. 
Even when $N$ is small, the Café~$\theta$ model successfully captures the overall trend of the distributions. 
However, there are cases where the model fails to reproduce the detailed shape of the distribution. 
For instance, some distributions exhibit peaks both at the center and the edges (as seen in Fig.~\ref{fig.05_a}), or show multimodal structures (e.g., in Fig.~\ref{fig.05_b} for $\alpha = 1/3200$ and $\theta = 0.95$), which are not accurately captured by the model. 

\subsection{Properties of Silence}

The following analysis focuses on the conditions under which silence arises in the Café~$\theta$ model, along with the expected duration of such silent periods. 

\vspace{2mm}
We begin by recalling the meanings of $\alpha$ and $\beta$: these parameters represent the transition probabilities from silence to speaking and from speaking to silence, respectively. 
As a result, increasing $\alpha$ shifts the overall distribution toward speaking, while increasing $\beta$ shifts it toward silence. 

\vspace{2mm}
However, our current interest lies not in such trivial tendencies, but in the behavior of silence under conditions where a non-negligible amount of speaking activity is present in the system. 
Specifically, we consider the case where the expected number of speaking individuals, $\langle n_1 \rangle$, is equal to $N/2$.

\vspace{2mm}
From Eq.~(\ref{eq.f_theta}), the expected value $\langle n_1 \rangle$ is given by: 

\begin{align*}
  \langle n_1 \rangle = N\frac{h_\alpha}{h_\alpha + h_\beta} = N\frac{\alpha}{\alpha + \beta}
\end{align*}

\vspace{2mm}
which is independent of $\theta$. 
It follows that the condition $\langle n_1 \rangle = N/2$ holds when $\alpha = \beta$.

\begin{figure}[htbp]
  \centering
  \includegraphics[width=1.0\hsize]{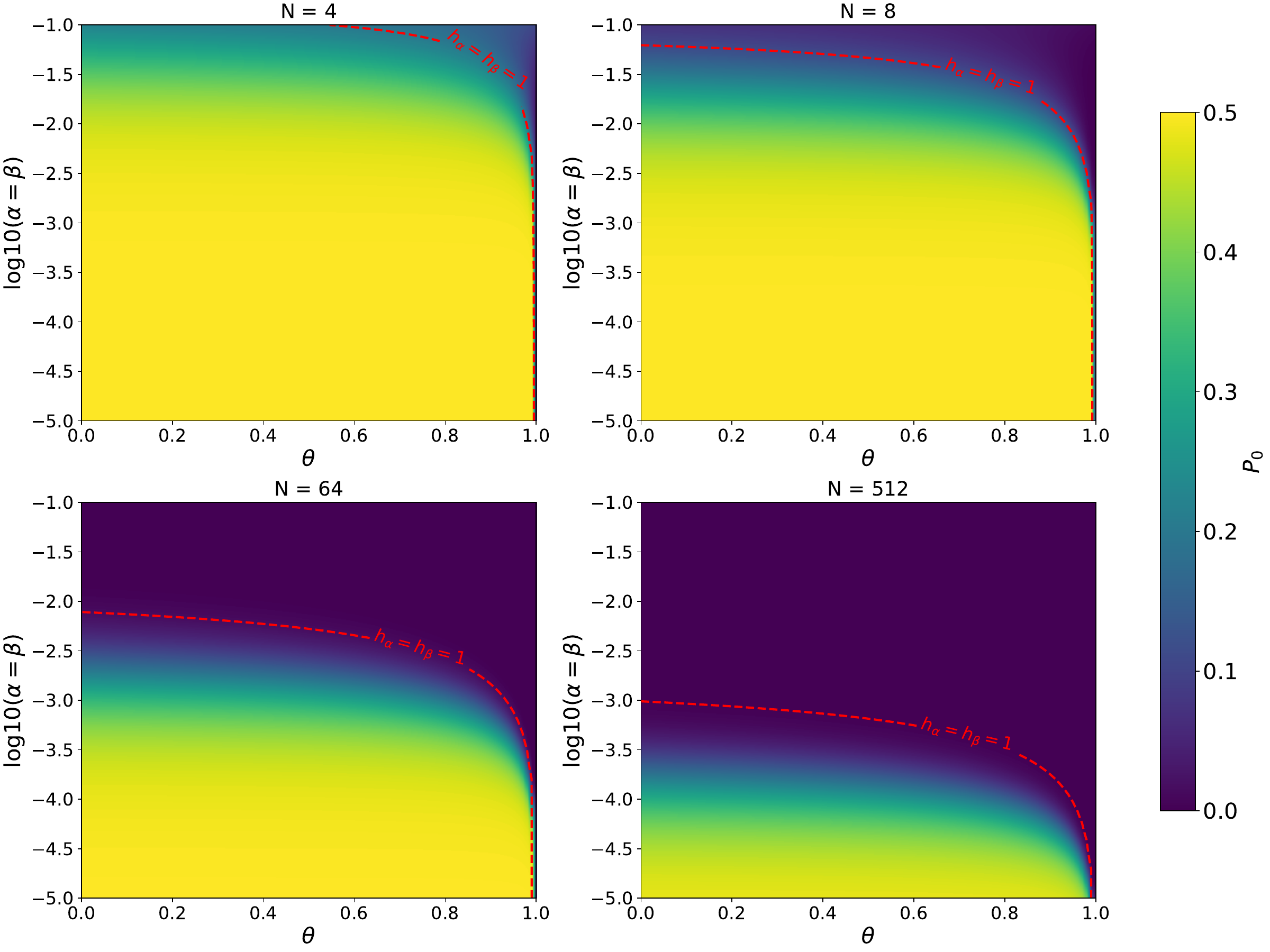}
  \caption{
  Heatmaps showing the silence probability $P_0 = f_\theta(n_1 = 0)$ for $N = 4, 8, 64, 512$, under varying values of $\alpha$, $\beta$, and $\theta$. 
  The dashed line indicates the region where $h_\alpha = h_\beta = 1$, corresponding to a uniform distribution of $f_\theta$. 
  The behavior of $P_0$ changes dramatically across this boundary. 
  }
  \label{fig.06}
\end{figure}

\vspace{2mm}
Under this condition, we consider the probability that silence occurs, i.e.,~the probability that no one is speaking, $f_\theta(n_1 = 0) = P_0$. 
For several values of $N$, we varied the parameters $\alpha$, $\beta$, and $\theta$, and plotted the resulting values of $P_0$ as heatmaps in Fig.~\ref{fig.06}.

\vspace{2mm}
The results show that $P_0$ tends to be high when both $\alpha$ and $\beta$ are small. 
This implies that silence is more likely to occur when the system has a strong tendency to remain in its current state. 
Furthermore, as $N$ increases, the region where $P_0$ is nearly zero becomes larger. 
This aligns with the intuitive idea that silence becomes less likely in larger groups. 

\vspace{2mm}
Next, we focus on the role of $\theta$. 
As $\theta$ increases, $P_0$ sharply drops to zero when $h_\alpha = h_\beta = 1$. 
This phase-transition-like behavior occurs precisely at the point where $f_\theta$ becomes a uniform distribution, and the qualitative nature of $f_\theta$ changes drastically across this threshold. 
In the region where $h_\alpha = h_\beta < 1$, the distribution $f_\theta(n_1)$ tends to concentrate near both extremes, $n_1 = 0$ and $n_1 = N$. This corresponds to a dynamical regime in which the system alternates between states where almost everyone is speaking and where almost everyone is silent. 
By contrast, in the region where $h_\alpha = h_\beta > 1$, $f_\theta$ resembles a normal distribution, and $P_0$ is close to zero. 
In other words, $\theta$ represents the degree to which an agent's own state influences its future behavior. When $\theta$ is large, agents become less responsive to their environment, and silence is less likely to occur. 
These results suggest that silence in a group setting can only emerge when each individual maintains at least some awareness of others. 

\vspace{2mm}
Finally, we consider the expected duration of a silent period, denoted $\tau$. 

The probability that at least one person begins to speak in the next time step, starting from a fully silent state, is given by the complementary probability: $1-(1-\alpha)^N$. This value does not depend on $\theta$. 
From this, the expected duration of silence is given by: 

\begin{align*}
  \langle \tau \rangle = \frac{1}{1-(1-\alpha)^N} \sim \frac{1}{\alpha N}
\end{align*}

\vspace{2mm}
This result supports the intuitive notion that silence tends to last longer when the transition probability $\alpha$ is small or when the number of agents $N$ is small. 

\begin{figure}[htbp]
  \centering
  \includegraphics[width=1.0\hsize]{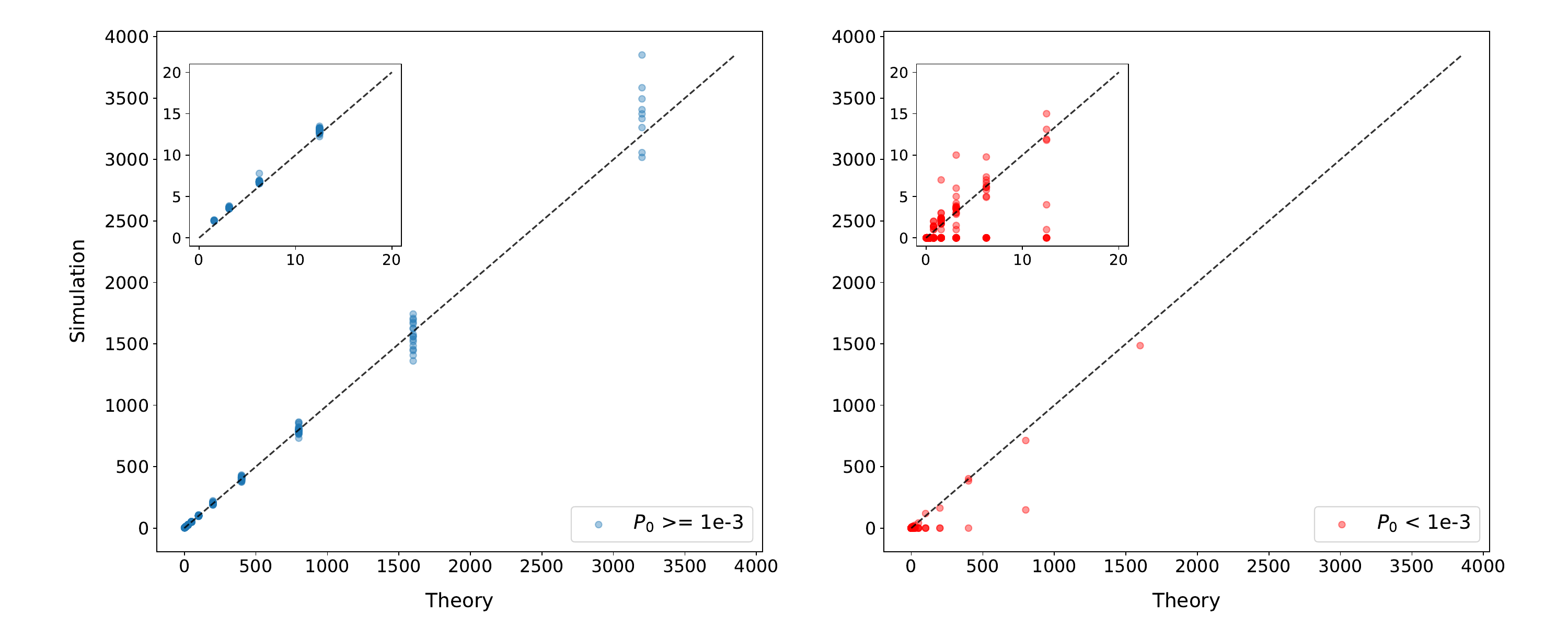}
  \caption{
  Comparison between the theoretical expected value and simulation results for the duration $\tau$ of silent periods. 
  The horizontal axis shows the theoretical value, and the vertical axis shows the average obtained from simulations. 
  The unit is time steps. 
  Since $\tau$ cannot be reliably measured when the silence probability $P_0$ is too low, the data is split into two plots at $P_0 = 10^{-3}$. 
  In the left plot, where $P_0$ is sufficiently high, the theoretical and simulated values show excellent agreement. 
  }
  \label{fig.09}
\end{figure}

\vspace{2mm}
Figure~\ref{fig.09} compares this theoretical expectation with the average duration observed in simulations. 
When the baseline probability of silence $P_0$ is low, it is difficult to estimate $\langle \tau \rangle$ accurately in simulations, and the measured value tends to be zero. 
We divided the plot at $P_0 = 10^{-3}$ to distinguish between high- and low-silence regimes. 
In the left plot, where $P_0$ is sufficiently large, the theoretical and empirical values of $\langle \tau \rangle$ agree closely. 
In this region, we also observe that the variance in $\tau$ increases along with its mean, which can be understood from the theoretical expression for the variance: 

\begin{align*}
  \mathrm{ Var } (\tau) = \frac{(1-\alpha)^N}{\left(1-(1-\alpha)^N\right)^2} \sim \frac{(1-\alpha)^N}{\alpha^2 N^2}
\end{align*}

\vspace{2mm}

\subsection{Comparison with Experimental Data}

We compare the stationary distribution of the Café~$\theta$ model with empirical data from the CHiME-6 dataset~\cite{2020WAT}. 
This dataset was designed for tasks such as speech separation and recognition in multi-speaker environments, and contains audio recordings of four-person dinner parties held in real home settings. 

\vspace{2mm}
It is important to note that this comparison is not ideal, as the Café~$\theta$ model was developed under the assumption that the number of agents $N$ is sufficiently large. 
Nevertheless, the CHiME-6 dataset includes the most natural and multi-speaker conversational data available to us, making it a suitable starting point for evaluating the model. 
In the future, more appropriate validation should be conducted using experimental data involving larger groups. 

\vspace{2mm}
Due to these circumstances, some methodological considerations are necessary when comparing the Café~$\theta$ model with the CHiME-6 data. 
One of the first things we observe from examining the dataset is that all participants belong to a single, unified conversation group. 
In other words, there is no fragmentation of the conversation into subgroups. 
As a result, only one speaker typically talks at a given time. 
By contrast, the Café~$\theta$ model assumes an environment more akin to a literal café, where multiple conversations occur in parallel, and it is natural for several people to be speaking simultaneously. 
Therefore, in CHiME-6, the intensity of the acoustic environment reflects the volume of individual speakers, whereas in the Café~$\theta$ model, it reflects the number of speakers active at a given moment. 

\vspace{2mm}
To bridge this conceptual gap, we reinterpret the Café~$\theta$ model as follows: 
We treat a group of multiple agents as a single meta-agent. 
The individual components are referred to as ``sub-agents,'' and their collective behavior is interpreted as that of a ``parent agent.'' 
The number of sub-agents currently speaking is then treated as the effective voice intensity of the parent agent. 
Letting $M$ be the number of sub-agents per parent agent, the voice level of each parent agent can range from 0 to $M$. 
In our analysis, we represent the four speakers in CHiME-6 as a total of $4M$ sub-agents in the Café~$\theta$ model. 
We chose $M = 8$ in this study, though the results remain largely unchanged for any $M \geq 4$. 

\vspace{2mm}
The distribution for comparison with the Café~$\theta$ model was extracted from the CHiME-6 audio data using the following procedure. 
First, a short-time Fourier transform (STFT) was performed with a frame length of 512 samples and a hop length of 256 samples. The energy for each frame was calculated from the power spectrum. 
Frames with zero energy were excluded, as they likely represent gaps in the recording rather than actual silence. 
Next, the energies were converted to decibel scale, and extreme outliers were removed. 
The resulting distribution was used as the target for model comparison. 

\begin{figure}[htbp]
  \centering
  \includegraphics[width=1.0\hsize]{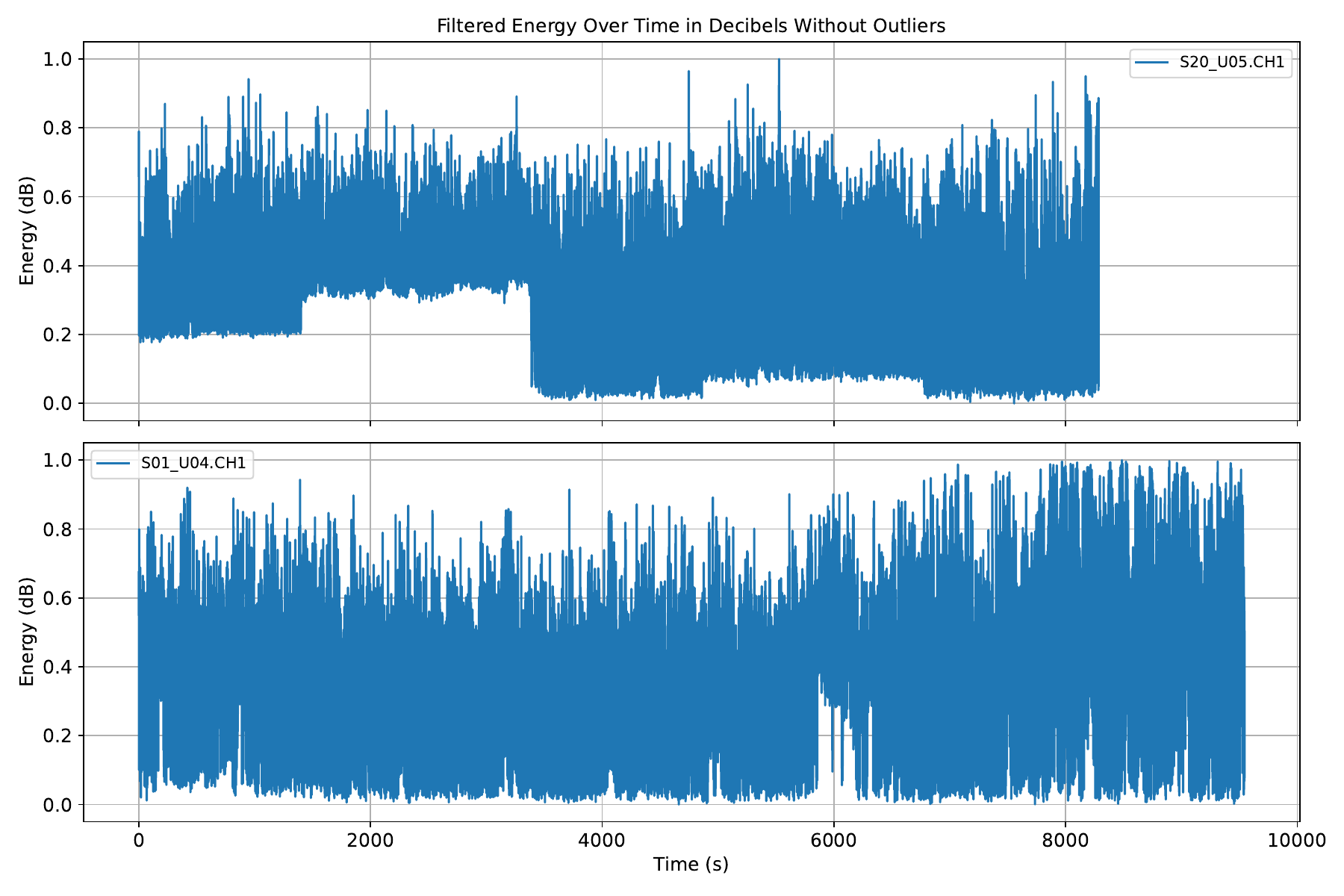}
  \caption{
  Temporal variation of frame-wise energy (in decibel scale) derived from the CHiME-6 power spectrum. 
  The top panel shows a case where the distribution changes significantly over time, while the bottom panel shows a case with a relatively stable distribution. 
  }
  \label{fig.08}
\end{figure}

\vspace{2mm}
CHiME-6 contains recordings from 24 sessions, each of which can be further divided into units based on recording location. 
Each session is approximately two hours long. Upon examining the distributions, we found that some recordings exhibit substantial shifts in distribution over time. 
Figure~\ref{fig.08} (top) shows an example of such a shift within a single session. 
These dynamic changes likely reflect transitions between qualitatively different types of conversation. 
In terms of the model, this could be interpreted as a change in parameters such as $\alpha$, $\beta$, or $w$. 
For this reason, we selected recordings that did not exhibit such distributional shifts. 
Specifically, each audio file was divided into 10 time segments, and the minimum decibel value was computed for each segment. We then selected recordings with the smallest variance among these minima. 
The example with the least variance is shown in the bottom panel of Fig.~\ref{fig.08}. 

\vspace{2mm}
We compared the Café~$\theta$ model to the nine recordings with the smallest variances. 
The model parameters were set to $h = 0$, with $\alpha$ and $\beta$ determined using the method of moments. 

\begin{figure}[htbp]
  \centering
  \includegraphics[width=1.0\hsize]{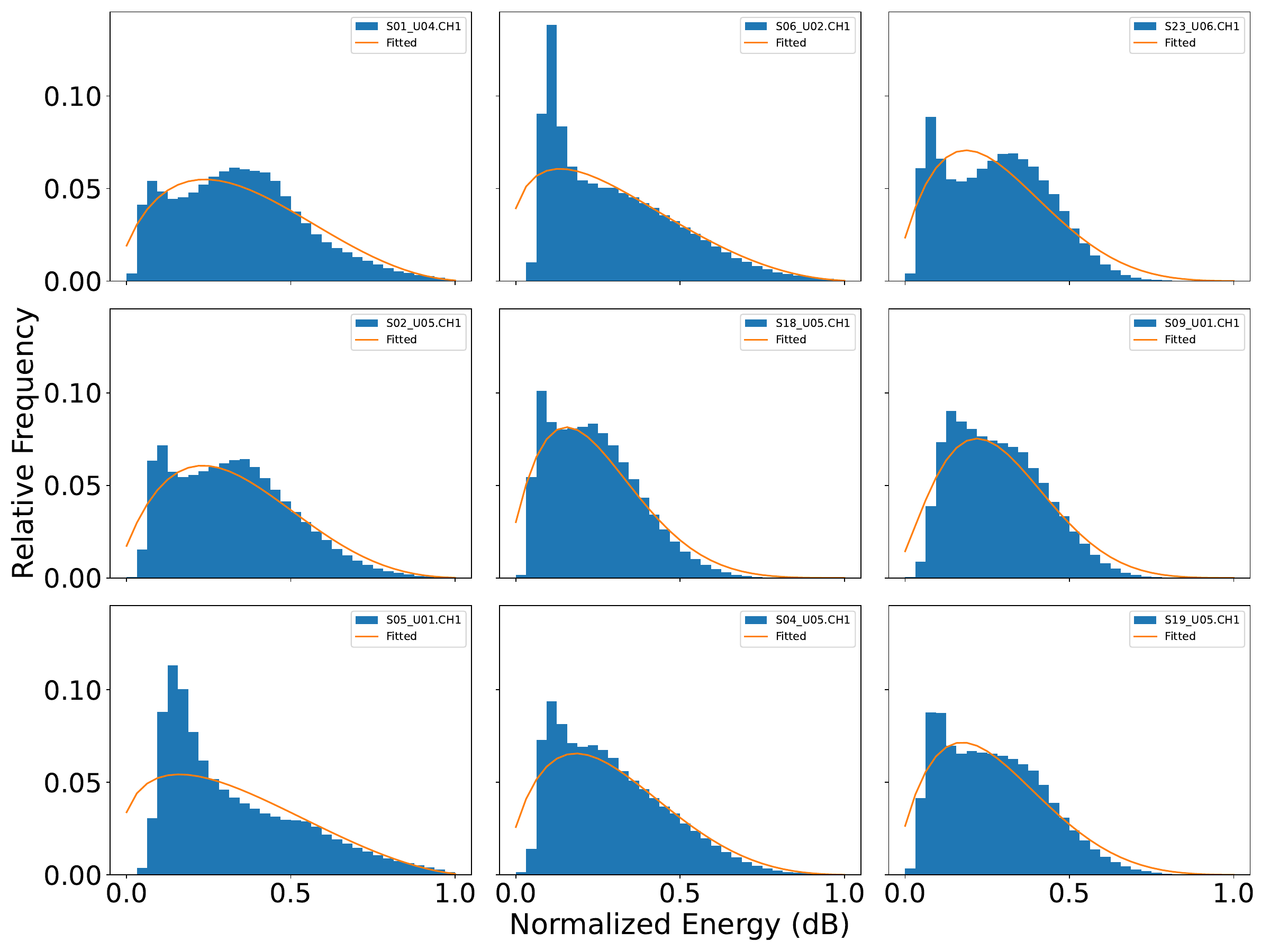}
  \caption{
  Comparison between CHiME-6 samples (with low variance) and the Café~$\theta$ model.  
  Although the model captures the general distributional shape, it fails to reproduce the multimodal structures present in some recordings.
  }
  \label{fig.07}
\end{figure}

\vspace{2mm}
As shown in Fig.~\ref{fig.07}, while the model captures the overall distributional trends reasonably well, several audio samples exhibit distinct multimodal features that cannot be reproduced by the Café~$\theta$ model’s simple beta-like distribution. 
Interestingly, this multimodality is consistent with what was previously observed in cases where the weight matrix $w$ was randomized (Fig.~\ref{fig.05}). 

\subsection{Challenges and Limitations of the Proposed Model}

In this section, we explore the limitations and challenges of the Markov chain model proposed in this study by reviewing key findings from the literature on the characteristics of real-life conversation. 

\vspace{2mm}
First, it is not difficult to imagine that the impact of speaking on listeners depends heavily on the content of the utterance, even when the act of speaking is the same. For example, seemingly meaningless responses such as ``uh huh'' can influence the listener’s level of engagement in the conversation~\cite{1982SCE}. 
Additionally, studies focusing on the formation of small conversational subgroups have shown that such splits are often triggered by specific utterances~\cite{2019RAM, 2023AOK2}. 
The model proposed in this paper, however, reduces all speaking behavior to a single uniform label (1) and thus cannot account for these nuanced effects. 

\vspace{2mm}
Furthermore, it has long been emphasized in conversation research that nonverbal communication plays a crucial role in shaping interaction~\cite{1959HAL, 1962AUS, 1988STR, 1996FOR, 1999CAR, 2003BON, 2003VER, 2004PEN, 2023AOK2, 2008STI, 2009STI, 2007BER, 2023KHK}. 
Participants constantly adjust their speaking based on mutual observation, and turn-taking is often guided by cues such as facial expressions, nodding, and gestures~\cite{1959HAL, 1996FOR, 2008STI, 2009STI, 2023KHK}. 
Physical proximity and body orientation also play an important role~\cite{2003BON}, and visual information can strongly influence the emergence of silence~\cite{2021ISH}. 
Despite this, all of these elements are entirely ignored in the current model. 
In particular, overlapping speaking—which is typically resolved quickly in natural conversation~\cite{2000SCH}—is instead emphasized in our model. 

\vspace{2mm}
Such discrepancies between the model and real-world interaction might be mitigated to some extent by extending the model. For instance, one could introduce finer-grained state classifications or allow for multiple types of interactions. 
Indeed, numerous studies have proposed typologies of silence~\cite{1995KUR, 2002AGY, 2007KUR, 2009BAC, 2012BRI, 2012STO, 2022HAO}, and attempts to classify different types of speaking date back many decades~\cite{1962AUS, 1982THO}. 
Allowing negative interaction terms, for example, could make it possible to model situations in which certain utterances suppress others—potentially improving the model’s ability to handle overlapping speaking and turn-taking behavior. 

\vspace{2mm}
That said, there are aspects of silence that remain difficult to capture even with such extensions. 
For example, research has shown that when a designated speaker delays their response, non-designated participants are more likely to attempt to take the turn~\cite{2006STI}. 
H. Maclay and C. E. Osgood analyzed English speech data recorded at a conference at the University of Illinois and identified four distinct types of hesitation in speaking~\cite{1959MAC}. 
Such hesitation has also been shown to influence the listener’s comprehension~\cite{1968MAR}. 
These subtle conversational dynamics are unlikely to be captured even by more elaborate extensions of the current model. 
Moreover, several studies suggest that silence within organizations is a collective phenomenon~\cite{2003FJM, 2011DON, 2012BRI, 2014MOR, 2016KNO, 2017WEI, 2022HAO, 2023MOR}. In such cases, silence often stems from factors such as awareness of organizational hierarchy or the perception that speaking up is futile—contexts far more complex than what the current model can handle. 
It is also important to note that these dynamics can be influenced by cultural and gender-related factors~\cite{2023MOR}. 

\vspace{2mm}
\section{Conclusion}
\label{sec.conc}

In this study, we analyzed the phenomenon of silence in co-present groups using a Markov chain model in which each agent can take one of two discrete states: silence ($0$) or speaking ($1$). 
We began by considering isolated agents with no mutual interactions, then examined the case of uniform interaction, and finally proposed the Café~$\theta$ model as an interpolation between these two extremes. 
Despite relying on several approximations, the Café~$\theta$ model showed excellent agreement with numerical simulations. Furthermore, the model captured the general shape of the distribution even under randomized interaction weights, indicating its robustness and flexibility. 
However, when the weights were random and the group size $N$ was small, the model failed to capture the observed multimodality. 
This highlights the limitations of the Café~$\theta$ model and suggests that the Markov model, beyond just the Café~$\theta$ variant, may be capable of expressing more complex distributions.

\vspace{2mm}
Using the Café~$\theta$ model, we investigated the conditions under which silence emerges. Our results showed that silence is more likely to occur when individuals have a strong tendency to maintain their current state or when the group size is small—findings that align with our intuition. 
Conversely, when individuals’ awareness of surrounding speaking is reduced, we observed a nontrivial, phase-transition-like behavior in which the probability of silence abruptly drops to zero beyond a certain threshold. 
To borrow from a French proverb, one might say that ``{\it Un ange passe}'' only when each person remains at least minimally attuned to those around them. 

\vspace{2mm}
We also conducted a preliminary comparison between the Café~$\theta$ model and the CHiME-6 dataset, which records four-person dinner party conversations. 
While we must reiterate that the Café~$\theta$ model is designed for larger groups, it was nevertheless able to reproduce the general distributional tendencies. 
However, similar to simulations with small group sizes, the CHiME-6 data exhibited multimodal structures that were not captured by the model, indicating the need for further investigation. 

\vspace{2mm}
Finally, we examined the characteristics of real-world conversation and discussed the limitations and challenges of the proposed model. 

\vspace{2mm}
In this study, we focused primarily on the static quantity of speaking distribution. 
However, the model also holds promise for analyzing conversational dynamics. 
The study of dynamic aspects of human communication has gained importance in the field of complexity science~\cite{2015SHE}. 
Future work may reveal the structure of conversational transitions by analyzing the dynamics before and after moments of silence using this model.
Additionally, the comparison with empirical data remains insufficient. 
In particular, to validate the Café~$\theta$ model rigorously, we require natural conversational data involving a sufficiently large number of individuals sharing the same physical space. 
Furthermore, it is necessary to discuss how to link experimental data with the model—specifically, how parameters such as $\alpha$, $\beta$, and $\theta$ should correspond to real-world phenomena. 
For example, a study by M.~Cristani {\it et al.} ~\cite{2010CRI} analyzed the temporal distributions of speaking and silence across various dialogue scenarios. Based on these findings, one could prepare different sets of $\alpha$ and $\beta$ values to investigate silence emergence conditions that reflect actual conversations. 
Alternatively, it may be possible to evaluate $\theta$—which reflects awareness of others’ speaking and is difficult to measure directly—by estimating $\alpha$ and $\beta$ from observable quantities such as average silence and speaking durations, and adjusting $\theta$ so that the model's speaking distribution aligns with experimental data. 

\vspace{2mm}
\section{Code Availability}
\label{sec.code}

The source code for the model discussed in this study is publicly accessible on GitHub \url{https://github.com/A5size/angel-passes} and is licensed under the terms of the MIT License. 

\vspace{2mm}
\section{Acknowledgments}

I would like to thank Kentaro Yonemura for his valuable feedback. 
I am also grateful to Yuta Yamashita for carefully reading the manuscript, identifying mistakes in my equations, and providing insightful comments. 
Finally, I am deeply grateful to Hirokazu Nitta for his generous support and encouragement. 

\vspace{2mm}

During the preparation of this manuscript, the authors used ChatGPT (GPT‑4o) exclusively to improve the clarity and readability of the English text throughout the paper. After using this tool, the authors reviewed and edited the output as necessary and accept full responsibility for the final content.

\appendix

\vspace{2mm}
\section{Derivation of the Uniform Group Distribution}
\label{sec.app1}

In the case where $w_{ij} = 1/N$, the probability $f_\mathrm{uni}(n_1)$ that $n_1$ agents are speaking after a sufficiently long time satisfies the following equation: 

\begin{align*}
  f_\mathrm{uni}(n_1') = \sum_{n_1=0}^{N} g(n_1, n_1') f_\mathrm{uni}(n_1)
\end{align*}

\vspace{2mm}
where $g(n_1, n_1')$ denotes the transition probability from $n_1$ to $n_1'$ speakers, and is given by:

\begin{align*}
  g(n_1, n_1') = 
  \binom{N}{n_1'}
  \left( \frac{n_0}{N}\alpha + \frac{n_1}{N}\left(1-\beta\right) \right)^{n_1'} 
  \left( \frac{n_0}{N}\left(1-\alpha\right) + \frac{n_1}{N}\beta \right)^{N-n_1'}
\end{align*}


\vspace{2mm}
In general, solving this equation analytically is difficult. 

Therefore, we approximate the solution by reformulating it as an integral equation. Assuming a functional form that includes unknown parameters, we then determine plausible values for those parameters. 

\vspace{2mm}
Let us define the normalized variables: 

\begin{align*}
  x &= \frac{n_1}{N}\\
  x' &= \frac{n_1'}{N}
\end{align*}

\vspace{2mm}
Using these, the original equation can be rewritten as the following integral equation: 

\begin{align*}
  \tilde{f}_\mathrm{uni}(x') &= 
  \frac{\Gamma\left(N+1\right)}{\Gamma\left(x'N+1\right)\Gamma\left(N-x'N+1\right)}\\
  &\qquad \times
  N\int_{0}^{1}
  \left\{ \left(1-x\right)\alpha + x\left(1-\beta\right) \right\}^{x'N}  
  \left\{ \left(1-x\right)\left(1-\alpha\right) + x\beta \right\}^{N-x'N}
  \tilde{f}_\mathrm{uni}(x)
  dx
\end{align*}


\vspace{2mm}
Now, let us define: 

\begin{align*}
  y &= \left(1-x\right)\alpha + x\left(1-\beta\right) \\
  y' &= \left(1-x'\right)\alpha + x'\left(1-\beta\right) 
\end{align*}

\vspace{2mm}
Then the integral equation becomes: 

\begin{align*}
  \tilde{f}_\mathrm{uni}(y')
  = 
  \frac{\Gamma\left(N+1\right)}{\Gamma\left(x'N+1\right)\Gamma\left(N-x'N+1\right)}
  \frac{N}{1-\alpha-\beta}
  \int_{\alpha}^{1-\beta}
  y^{x'N} \left( 1 - y \right)^{N-x'N}
  \tilde{f}_\mathrm{uni}(y)
  dy
\end{align*}


\vspace{2mm}
Assuming that $\tilde{f}_\mathrm{uni}(x)$ can be written using the beta distribution with parameters $a$ and $b$: 

\begin{align*}
  \tilde{f}(x) = \frac{1}{\mathrm{B}(a, b)} y^{a-1} \left( 1 - y \right)^{b-1} 
\end{align*}

\vspace{2mm}
we obtain: 

\begin{align*}
  y'^{a-1} \left( 1 - y' \right)^{b-1} =
  \frac{\Gamma\left(N+1\right)}{\Gamma\left(x'N+1\right)\Gamma\left(N-x'N+1\right)}
  \frac{N}{1-\alpha-\beta}
  \int_{\alpha}^{1-\beta}
  y^{a + x'N - 1} \left( 1 - y \right)^{b + N-x'N -1}
  dy
\end{align*}

\vspace{2mm}
Since $0 < \alpha, \beta \ll 1$, we approximate the integration range from $[ \alpha, 1 - \beta ]$ to $[0, 1]$, yielding: 

\begin{align*}
  &\int_{\alpha}^{1-\beta}
  y^{a + x'N - 1} \left( 1 - y \right)^{b + N-x'N -1}
  dy\\
  &\sim
  \int_{0}^{1}
  y^{a + x'N - 1} \left( 1 - y \right)^{b + N-x'N -1}
  dy
  =
  \frac{\Gamma\left(a+x'N\right)\Gamma\left(b+N-x'N\right)}{\Gamma\left(a+b+N\right)}
\end{align*}


\vspace{2mm}
Thus, we arrive at the following equation for determining the parameters $a$ and $b$: 

\begin{align*}
  y'^{a-1} \left( 1 - y' \right)^{b-1} =
  \frac{\Gamma\left(N+1\right)}{\Gamma\left(x'N+1\right)\Gamma\left(N-x'N+1\right)}
  \frac{N}{1-\alpha-\beta}
  \frac{\Gamma\left(a+x'N\right)\Gamma\left(b+N-x'N\right)}{\Gamma\left(a+b+N\right)}
\end{align*}


\vspace{2mm}
We determine $a$ and $b$ so that both sides are approximately equal at $x' = 1/2$. 
This is justified because, for sufficiently large $z$, Stirling's approximation

\begin{align*}
  \Gamma (z)\sim {\sqrt {\frac {2\pi }{z}}}\left({\frac {z}{e}}\right)^{z}
\end{align*}

\vspace{2mm}
can be applied. 

\vspace{2mm}
Replacing the gamma functions yields: 

\begin{align*}
  &\left\{ \left(1-x'\right)\alpha + x'\left(1-\beta\right) \right\}^{a-1}  
  \left\{ \left(1-x'\right)\left(1-\alpha\right) + x'\beta \right\}^{b-1}\\
  &\qquad =
  \frac{\sqrt{N}}{1-\alpha-\beta}
  \frac{(a + x'N)^{a + x'N - \frac{1}{2}} (b + N - x'N)^{b + N - x'N - \frac{1}{2}}}
  {x'^{x'N + \frac{1}{2}}(1 - x')^{N - x'N + \frac{1}{2}}(a + b + N)^{a + b + N - \frac{1}{2}}}
\end{align*}

\vspace{2mm}
Substituting $x' = 1/2$, we obtain: 

\begin{align*}
  \left(1+\alpha-\beta\right)^{a-1}\left(1+\beta-\alpha\right)^{b-1} =
  \frac{1}{1-\alpha-\beta}
  \frac{\left(\frac{2a}{N}+1\right)^{a+\frac{N-1}{2}}\left(\frac{2b}{N}+1\right)^{b+\frac{N-1}{2}}}
  {\left(\frac{a+b}{N}+1\right)^{a+b+N-\frac{1}{2}}}
\end{align*}


\vspace{2mm}
Introducing a new parameter $c$ such that $a = c\alpha N$ and $b = c\beta N$, the equation becomes: 

\begin{align*}
  \left(1+\alpha-\beta\right)^{c\alpha N-1}\left(1+\beta-\alpha\right)^{c\beta N-1} =
  \frac{1}{1-\alpha-\beta}
  \frac{\left(2c\alpha+1\right)^{c\alpha N+\frac{N-1}{2}}\left(2c\beta+1\right)^{c\beta N+\frac{N-1}{2}}}
  {\left(c\alpha+c\beta+1\right)^{c\alpha N+c\beta N+N-\frac{1}{2}}}
\end{align*}


\vspace{2mm}
Taking logarithms of both sides, neglecting higher-order terms of $\alpha$ and $\beta$, and using the approximation $\log(1 + x) \sim x$, we obtain $c = 2$. 

\vspace{2mm}
Consequently, $\tilde{f}_\mathrm{uni}(x)$ is given by: 

\begin{align*}
  \tilde{f}_\mathrm{uni}(x) = 
  \frac{1}{\mathrm{B}(2\alpha N, 2\beta N)} 
  \left\{ \left(1-x\right)\alpha + x\left(1-\beta\right) \right\}^{2\alpha N-1}  
  \left\{ \left(1-x\right)\left(1-\alpha\right) + x\beta \right\}^{2\beta N-1}
\end{align*}


\vspace{2mm}
Since we treated $x = n_1/N$ as a continuous variable, applying this expression directly to discrete values causes the total probability to deviate from unity. 
Therefore, for the discrete case, we must re-normalize using the following definition: 

\begin{align*}
  f_\mathrm{uni}(n_1) = 
  \frac{1}{\mathrm{Z}(2\alpha N, 2\beta N)} 
  \left\{ \left(1-\frac{n_1}{N}\right)\alpha + \frac{n_1}{N}\left(1-\beta\right) \right\}^{2\alpha N-1}  
  \left\{ \left(1-\frac{n_1}{N}\right)\left(1-\alpha\right) + \frac{n_1}{N}\beta \right\}^{2\beta N-1}
\end{align*}


\vspace{2mm}
where the normalization constant $Z$ is given by: 

\begin{align*}
  \mathrm{Z}(2\alpha N, 2\beta N) = 
  \sum_{n_1=0}^{N} 
  \left\{ \left(1-\frac{n_1}{N}\right)\alpha + \frac{n_1}{N}\left(1-\beta\right) \right\}^{2\alpha N-1}  
  \left\{ \left(1-\frac{n_1}{N}\right)\left(1-\alpha\right) + \frac{n_1}{N}\beta \right\}^{2\beta N-1}
\end{align*}

\vspace{2mm}
For large $N$, this sum can be approximated as: 

\begin{align*}
  \mathrm{Z}(2\alpha N, 2\beta N) \sim \frac{N}{1-\alpha-\beta} \mathrm{B}(2\alpha N, 2\beta N)
\end{align*}


\vspace{2mm}
Alternatively, when $N$ is not very large, a trapezoidal approximation yields: 

\begin{align*}
  \mathrm{Z}(2\alpha N, 2\beta N) \sim 
  \frac{N}{1-\alpha-\beta} \mathrm{B}(2\alpha N, 2\beta N)
  + \frac{1}{2}
  \left( 
    \alpha^{2\alpha N - 1} (1-\alpha)^{2\beta N - 1} + (1-\beta)^{2\alpha N - 1} \beta^{2\beta N - 1}  
  \right)
\end{align*}


\bibliographystyle{unsrtnat}
\bibliography{references.bib}
\end{document}